\documentclass[aps,prd,twocolumn,showpacs,superscriptaddress,nofootinbib]{revtex4-2}

\usepackage{graphicx}
\usepackage{dcolumn}
\usepackage{bm}

\usepackage{overpic}
\usepackage[english]{babel}
\usepackage{graphicx}
\usepackage{epsfig}
\usepackage{amssymb}
\usepackage{amsmath}
\usepackage[plainpages=false,pagebackref=false,pdftex]{hyperref}
\usepackage{ulem}
\usepackage[T1]{fontenc} 
\usepackage[utf8]{inputenc}
\usepackage{color}

\usepackage[titletoc]{appendix}

\begin{document}
\title{Thermalization and prethermalization
in the soft-wall AdS/QCD model}

\author{Xuanmin Cao}
\affiliation{Department of Physics and Siyuan Laboratory, Jinan University, Guangzhou 510632, China}
\author{Jingyi Chao}
\email{Co-corresponding author: chaojingyi@jxnu.edu.cn}
\affiliation{ College of Physics and Communication Electronics, Jiangxi Normal University, Nanchang, Jiangxi 330022, China}
\author{Hui Liu}
\affiliation{Department of Physics and Siyuan Laboratory, Jinan University, Guangzhou 510632, China}
\author{Danning Li}
 \vspace{1cm}
\email{Corresponding author: lidanning@jnu.edu.cn}
\affiliation{Department of Physics and Siyuan Laboratory, Jinan University, Guangzhou 510632, China}




\begin{abstract}
The real-time dynamics of chiral phase transition is investigated in a two-flavor ($N_f=2$) soft-wall AdS/QCD model. To understand the dynamics of thermalization, we quench the system from initial states deviating from the equilibrium states. Then, we solve the nonequilibrium evolution of the order parameter (chiral condensate $\langle \sigma\equiv\bar{q}q\rangle$). It is shown that the system undergoes an exponential relaxation at temperatures away from the critical temperature $T_c$. The relaxation time diverges at $T_c$, presenting a typical behavior of critical slowing down. Numerically, we extract the dynamic critical exponent $z$, and get $z\approx 2$ by fitting the scaling behavior $\sigma\propto t^{-\beta/(\nu z)}$, where the mean-field static critical exponents (order parameter critical exponent $\beta=1/2$, correlation length critical exponent $\nu=1/2$ ) have been applied. More interestingly, it is remarked that, for a large class of initial states, the system would linger over a quasi-steady state for a certain period of time before the thermalization. It is suggested that the interesting phenomenon, known as prethermalization, has been observed in the framework of holographic models. In such prethermal stage, we verify that the system is characterized by a universal dynamical scaling law and described by the initial-slip exponent $\theta=0$.
\end{abstract}


\maketitle


\section{Introduction}\label{intro}
The nonequilibrium dynamics plays an essential role from the high energy physics and cosmology to condensed matter physics, at which relevant physical properties exhibit universal dynamic scaling behavior. The observed phenomena in relativistic heavy-ion collisions~\cite{BRAHMS:2004adc} lead to a consistent framework of bulk QCD matter evolutions and critical fluctuations which has become significantly important to address the critical slowing down near a critical point (CP)~\cite{Berdnikov:1999ph}, and to reveal the domain formation at the first-order phase transition~\cite{Randrup:2010ax}. Recently, the precision cosmology has firmly supported the Big Bang paradigm~\cite{LIGOScientific:2016aoc}. The description of the cosmological fluctuations and the subsequent dynamical response are in particular required to establish the scenarios of the baryon asymmetry and the dark matter production~\cite{Barnaby:2009mc}. In quantum systems of ultracold atomic gases~\cite{RevModPhys.80.885}, eventually, many out of equilibrium progresses are set into the context of universal dynamics in recent decades~\cite{ueda2020quantum}.

Among very different physical systems, a universal time evolution appearing in the early-time regime was discovered by J. Berges \textit{et al}. in Ref.~\cite{Berges:2004ce} through a quark-meson model. It is found that, governed by the strong fields or large occupancy of modes, a transient occurs in the far-from-equilibrium initial conditions, named as prethermalization. The observed evolution suggests that the systems are usually passing in the vicinity of a non-thermal fixed point (NTFP) before approaching to the long-time thermalization~\cite{Berges:2004ce,Berges:2020fwq}. In regard to the numerical value of the scaling exponents, one observes that several macroscopic properties of the underlying system play a vital role, such as the number of spatial dimensions, and whether the particle or energy cascade is being produced. Many relevant works of prethermalization have been completed within a variety of many-body models after sudden quenches of the phase transition parameters. In general, they can be classified into isolated systems~\cite{Prufer:2018hto,Erne:2018gmz,PineiroOrioli:2015cpb,Chiocchetta:2016rdr,Berges:2008wm,Berges:2016nru}  and open systems~\cite{Bonart:2012cri,PhysRevLett.113.220401}.  In a few integrable isolated systems, the prethermal states turn out to be described by the generalized Gibbs ensemble (GGE)~\cite{PhysRevLett.98.050405,PhysRevB.84.054304,Takashi:2018the}, though a general conclusion for any system is not reached yet. Under the renormalization group (RG) flow frame, the behaviors near the NTFP have been discussed~\cite{Chiocchetta:2016rdr,Berges:2008wm,Berges:2016nru}.

On the other hand, prior to reaching equilibrium, another possible pre-equilibrium critical phenomena emerges as the memory of the initial condition during the early stage and instead has been discovered earlier by Janssen \textit{et al.}~\cite{janssen1989new}. In the regime near the continuous phase transition, the initial preparation extends to all times in a manner similar to the surface critical phenomena. To describe the distribution width of initial configurations, the universal initial-slip exponent $\theta$ is proposed in the pioneering work of~\cite{janssen1989new} with a pure dissipative classical system. Here, the universal collective behavior is developed in the time window between the microscopic time scales and the translationally invariant asymptotics. Critical exponent $\theta$ specifically characterizes the breaking of the time translation invariance due to the initial conditions. The study of short-time scaling has already attracted much attention in the toy model of $\phi^4$ theory through the nonequilibrium renormalization group method~\cite{janssen1989new, Schoeller:2009ape, Berges:2000ew} and by the large $N$ expansion~\cite{Gagel:2015opa}. In many other open systems, further investigations on the short-time dynamical scaling are currently under active exploration~\cite{sieberer2016keldysh}.

Whiles the unusual initial states are created in the off-central heavy ion collisions, we expect that a potentially unique signature of magnetic fields would be manifested in the out-of-equilibrium evolution near the critical end point (CEP). Generally, an ultra-strong magnetic field is produced by the fast colliding, highly charged nuclei in heavy ion experiments. And the strength of the $eB$-field is at the order of $m_{\pi}^{2}$~\cite{Deng:2012pc}. However, the duration of the magnetic field remains an open question. A reasonable estimation is that the life-time of the $eB$-field is as long as the starting time of the hydrodynamic evolution, $\sim 0.6\,\text{fm}$~\cite{Xu:2020sui}. The rapidly decayed magnetic field leads to an undetectable signal of the chiral magnetic effect~\cite{Kharzeev:2013ffa}. As it turned out, such a quickly disappeared field could be served as a randomly prepared initial state for the evolution of the hydrodynamic quark-gluon plasma. And we expect that the short-time scaling is nevertheless encoded in the evolutions of two- and higher-point correlation functions~\cite{Stephanov:2011pb}. The influence of the initial slip exponent in the context of heavy ion collisions is not yet completed, in contrast with the relativistic hydrodynamics~\cite{An:2021wof}. This is because that a full theory which incorporate the fluctuations near the phase transition and bulk evolution in strongly coupling plasma is currently under construction.

As a powerful tool for addressing strongly-coupled gauge theories, the holographic duality has successfully predicted a lower bound of the shear viscosity over the volume density of entropy $\eta/S\leq 1/4\pi$~\cite{PhysRevLett.87.081601,PhysRevLett.93.090602,PhysRevLett.94.111601}. Impressively, the holographic method not only takes the advantage of addressing the near equilibrium phenomena for strongly coupling systems in a perturbative manner, but also is adept at studying the far-from-equilibrium dynamic processes. For the out-of-equilibrium dynamics, it is proposed that the linear relaxation is corresponding to the  quasinormal mode (QNM) of the black hole in the AdS/CFT correspondence~\cite{Kovtun:2005qua}. The nonequilibrium dynamics in the holographic model actually is the problem in classical general relativity, which can be solved with numerical relativity. The holographic duality has been successfully used to study the far-from-equilibrium dynamic phenomena, for example, the Kibble-Zurek mechanics in the holographic superfluid or superconductivity~\cite{PhysRevX.5.021015,Sonner:2014tca,PhysRevD.101.026003,Liu:2018crr}, the NTFP in the holographic superfluid~\cite{Ewerz:2014tua},  the holographic thermalization of super Yang-mills theory and QCD~\cite{PhysRevLett.102.211601,PhysRevD.82.026006,Rajagopal:2016uip,Atashi:2016fai,Casalderrey-Solana:2013aba,Critelli:2017euk,Ishii:2015gia}.

In this work, we will study the nonequilibrium physics of  QCD matter in the holographic framework, focusing on both the long time thermalization and the short time scaling behavior. Therefore, a holographic description of both the light modes and phase transitions would be quite necessary. In the bottom-up approach, the soft-wall AdS/QCD model proposed in Ref.~\cite{Karch:2006pv} does provide an effective scenario to consider light meson spectrum ~\cite{Kelley:2010mu,Sui:2009xe,Colangelo:2008us,Ballon-Bayona:2020qpq,FolcoCapossoli:2019imm,Li:2012ay,Li:2013oda,Cao:2021tcr, Cao:2020ryx,Rinaldi:2022dyh,Afonin:2021cwo,Mamedov:2021dpv,Chen:2021wzj,Capossoli:2021ope}, the chiral phase transition~\cite{Colangelo:2011sr,Chelabi:2015gpc,Li:2016smq,Chelabi:2015cwn,Fang:2018axm,Rodrigues:2020ndy,Bartz:2016ufc,Bartz:2017jku,Fang:2016nfj,Li:2016gfn}, as well as pion condensed phase~\cite{Lv:2018wfq,Cao:2020ske}. Thus, we will take the soft-wall model as our start point.

In the following, we will introduce the soft-wall AdS/QCD model and  holographic chiral phase transition in Sec.~\ref{model}. Then, in Sec.~\ref{sec:thermalization}, we will verify the linear relaxation and the critical slowing down with the time-dependent sigma condensate in the soft-wall AdS/QCD model, and the relations between the thermalization and the QNM. In Sec.~\ref{sec:prethermalization}, we will study the prethermalization as well as the crossover to the thermalization through different quench protocols. The short-time dynamic exponent is numerical fitted and $\theta=0$. Finally, a conclusion and discussion are given in Sec.~\ref{sec:summary}.

\section{The soft-wall AdS/QCD model and chiral phase transition}\label{model}
In this section, we will briefly review the soft-wall AdS/QCD model, which is a bottom-up holographic QCD model, based on the global ${\rm{SU}}(N_f)_L\times {\rm{SU}}(N_f)_R$ chiral flavor symmetry. The original soft-wall model is proposed by Karch {\it{et al.}} in Ref.~\cite{Karch:2006pv}. As mentioned above, by slightly extending the original soft-wall model, the chiral phase transition and light meson spectra could be well described. Furthermore, the mass diagram of chiral phase transition could be realized~\cite{Chen:2018msc}, in which a critical point appear in two-flavor chiral limit. Thus, it provides an ideal start point for the main interests of this work.

In the soft-wall models, the background metric is usually taken as the anti-de Sitter (AdS) metric
\begin{eqnarray}
ds^2=&e^{2A(r)}(-dr^2+\eta_{\mu\nu}dx^\mu dx^\nu),
\end{eqnarray}
where $A(r)=-\ln r$, $\eta_{\mu\nu}={\rm{diag}}(1,-1,-1,-1)$,  $x^{\mu}$ represents the 4D time and space dimensions, and $r$ represents the fifth dimension.
In this work, we concentrate on bulk scalar field part and neglect the gauge field part of the 5D action. Thus, the 5D action in the bulk is
\begin{eqnarray}\label{action-SW}
S=\int d^5 x e^{-\Phi (r)}\sqrt{g}\    {\rm{Tr}}\left\{|DX|^2-(m_5^2 |X|^2+\lambda |X|^4)\right\},\nonumber\\
\end{eqnarray}
in which $X$ is the bi-fundamental scalar field;  $m_5$ is the 5D mass of $X$; $\sqrt{g}$ is the determinant of the metric and $\lambda$ is a fitting parameter of the quartic term of the potential. $\Phi (r)$ is the dilaton field which is introduced as a smooth cutoff, and it is very essential for the Regge-like behavior of the mass spectrum. To be simplicity, we employ the degenerate $N_f=2$ case with $m_u=m_d=m_q$, thus $X$ can be defined as
\begin{align}
    X=\begin{bmatrix}
    \frac{\chi+S}{2}&0\\ 0& \frac{\chi+S}{2}
    \end{bmatrix} {\rm{Exp}}{[2 i \pi^i t^i]},
\end{align}
where $t^i$ are the generators of ${\rm{SU}}(2)$; $S$ is the scalar perturbation and $\pi^i$ are the pseudo-scalar perturbations. The near boundary ($r=0$) expansion of the bulk scalar field $\chi(r)$ could be derived from the equation of motion as
\begin{align}\label{eq:boundaryofchi}
    \chi(r\rightarrow 0)=m_q \gamma r+...+\frac{\sigma}{\gamma}r^3+...,
\end{align}
with  the chiral condensate $\sigma=\langle \bar{q}q\rangle$. The parameter $\gamma=\sqrt{N_c}/2\pi$ (the number of colors $N_c=3$ ) is a normalization constant, which is fixed by matching the 4D two-point correlator~\cite{Cherman:2008eh}.

In thermal medium, one often takes the AdS-Schwarzchild black hole solution as the background metric
\begin{subequations}
\begin{eqnarray}
    ds^2&=&e^{2 A(r)}\left\{-\frac{1}{f(r)}dr^2+f(r)dt^2-d\vec x^2\right\},\\
    f(r)&=&1-\frac{r^4}{r_h^4},
\end{eqnarray}
\end{subequations}
with the position of the horizon at $r=r_h$ and $0\leq r\leq r_h$. From the holographic dictionary, one can identify the temperatrue $T$ with the Hawking temperature of the black hole
\begin{equation}
    T=\frac{1}{\pi r_h}.
\end{equation}
 As shown in Refs.~\cite{Chelabi:2015gpc,Chen:2018msc}, the original soft-wall model gives a vacuum without chiral condensate in the chiral limit $m_q=0$. Therefore, to reproduce the correct behavior, following Ref.~\cite{Fang:2016nfj}, we introduce a $r$ dependent form of the 5D mass $m_5^2$ and keep the quadratic dilaton, i.e., taking
\begin{subequations}
\begin{eqnarray}
m_5^2&=&-3-\mu_c^2 r^2,\label{eq-m52}\\
\Phi (r)&=&\mu_g^2 r^2,
\end{eqnarray}
\end{subequations}
where $\mu_c,\mu_g$ are  two model parameters. There could be several possible kinds of origins of this modification. One might consider it as coming from the anomalous dimension correction to the operator dimension $\Delta(r)$, the $r$ (or energy scale) dependence of which would lead to a $r$ depdent $m_5^2(r)$. It could also come from a coupling between $X$ and $\Phi$, representing the interaction of the flavor part with the background, and the mass term would effectively become $m_5^2\rightarrow{m_5^2+h(\Phi)}$ with $h$ a function of $\Phi$. As a phenomenological model, here we would not try to derive the exact form of the corrections. Instead, we will mainly focus on the qualitative behavior. So we just follow Ref.~\cite{Fang:2016nfj} and take the  simple form in Eq.\eqref{eq-m52}. From our numerical calculation, though the quantitative quantities (like the location of the critical point)  would depend on the value of $\mu_c$, the qualitative behavior discussed in this work would not be changed, when we change the value of $\mu_c$ guaranteeing the existence of the critical point. Actually, when we take another form of $m_5^2$ (like $m_5^2=-3+\kappa_1 \tanh(\kappa_2\Phi)$ , in another work of us in progress~\cite{comingwork}), or even when we take another form of the dilaton field as in Ref.\cite{Chelabi:2015gpc},  the qualitative results remain the same. Thus, the qualitative behavior discussed below would depend only on the existence of the critical point, and we will stick to the model in Eq.\eqref{eq-m52} in this work. The values of $\mu_c,\mu_g, m_q, \lambda$ are fitted by the hadron spectra. With $\mu_c=1450$ MeV, $\mu_g=440$ MeV, $m_q=3.22$ MeV and $\lambda=80$, it gives a physical pion mass $m_{\pi}=139.7$ MeV.

In this model, for any finite quark mass, the chiral phase transition is a crossover. To study the critical phenomena, one has to take the chiral limit $m_q=0$. In this limit, the crossover transition turns to a second order transition.\\
After fitting the model parameter by the hadronic spectra, we can study the chiral phase transition in the chiral limit and obtain the critical temperature $T_c\approx 0.16332301$ GeV and the saturation value of the sigma condensate $\sigma_{sat}\approx 0.015$ $\text{GeV}^3$~\cite{Cao:2020ryx}. It is interesting to see that it is comparable with the results of the critical temperature from lattice QCD and holographic models, like lattice results $T_c=171\pm 4$ MeV in Ref.~\cite{CP-PACS:2000phc}, $T_c=154\pm 9$ MeV in Ref.~\cite{PhysRevD.85.054503} and holographic model results $T_c=210$ MeV in Ref.~\cite{Colangelo:2011sr}, $T_c=151$ MeV in Ref.~\cite{Chelabi:2015gpc}.

In this work, we will study the universal properties of the nonequilibrium QCD in this IR-modified soft-wall AdS/QCD model.  In the soft-wall AdS/QCD model, the mesons are considered as the perturbations on the fixed background metric. In other words, we will consider a subsystem, which is coupled to an infinite large heating bath.

\section{Thermalization}\label{sec:thermalization}
In the previous studies, the scalar field $X$ is considered as time independent field to describe the thermal equilibrium state. To extend those studies to nonequilibrium cases, one has to study the time evolution of $X$. To avoid the divergence near the horizon, one would transform the coordinates to the Eddington-Finkelstein (EF) coordinates,
\begin{eqnarray}
t\rightarrow v&=&t-h(r),\\
h'(r)&=&\frac{1}{f(r)},\label{hprim}
\end{eqnarray}
with a new `time' coordinate $v$. By properly choosing the integral constant in Eq.\eqref{hprim}, one can set $t=v$ at the boundary $r=0$. Thus,
the AdS-Schwarzchild metric becomes
\begin{eqnarray}\label{eom-chi-v}
ds^2=e^{2A(r)}\{f(r) dv^2+2 dv dr-d\vec{x}^2\}.
\end{eqnarray}

Under the EF coordinate background metric, one obtains the equation of motion (EOM) of $\chi$ as~\footnote{We solve the EOM (Eq.~\ref{eom-chi}) through the pseudospectral method~\cite{boyd2001chebyshev,hesthaven2007spectral}. A brief introduce for the solving processes is given in Appendix~\ref{appendixA}.}
\begin{widetext}
\begin{eqnarray}\label{eom-chi}
2\partial_v\partial_r \chi(v,r)-\left[\frac{3}{r}+\Phi '(r)\right]\partial_v \chi (v, r)-f(r)\partial_r^2\chi(v,r)&\nonumber\\
+\left[\frac{3}{r} f(r)+\Phi'(r)f(r)-f'(r)\right]\partial_r\chi(v,r)+\frac{1}{ r^2}(m_5^2+\frac{\lambda}{2}\chi(v,r)^2)\chi (v,r)&=0.
\end{eqnarray}
\end{widetext}
As shown in Ref.~\cite{Chen:2018msc,Cao:2021tcr}, in the case of $N_f=2$, the second-order phase transition point (critical point) appears only in the chiral limit.  Therefore, we will focus on the two flavor case with zero quark mass for studying the phenomena in the critical region.

\vspace{1cm}
\subsection{Relaxation and critical slowing down}\label{sec:relaxation}
The static chiral phase transition is a second-order phase transition in the chiral limit, with the symmetry spontaneously breaking from ${\rm{SU}}(2)_L\times{\rm{SU}}(2)_R$ to ${\rm SU}_V(2)$. From a previous work, in Ref.~\cite{Chen:2018msc}, the chiral phase transition in the soft-wall model is very similar to the mean field  3D Ising model, whose critical exponents: the order parameter critical exponent $\beta=1/2$ , the correlation length critical exponent $\nu=1/2$, and the critical exponent of extra-field versus order parameter $\delta=3$. To extend to nonequilibrium physics, we
study the dynamic relaxation properties and the critical slowing down behavior described by this modified soft-wall AdS/QCD model in this section.

\subsubsection{Relaxation time}
Within the linear response theory~\cite{SUZUKI1976435}, either the perturbation taking place in the condensed phase (ordered phase) or in the chiral symmetry restored phase (disordered phase), the sigma condensate $\sigma(t)$ relaxes as
\begin{eqnarray}\label{evolvingEOMs}
\frac{\partial}{\partial t}\sigma(\epsilon,t)&=&-\frac{\sigma(\epsilon,t)-\sigma_{eq}}{\tau_R},
\end{eqnarray}
with the distance to the critical point $\epsilon=(T-T_c)$, the equilibrium sigma condensate $\sigma_{eq}=\sigma(\epsilon,\infty)$ and the relaxation time $\tau_R$. Then, one has $\sigma(\epsilon,t)$,
\begin{equation}\label{evolvingEOMs2}
\sigma(\epsilon,t)=\sigma_{eq}+\left [\sigma(\epsilon,t_0)-\sigma_{eq}\right ]e^{-(t-t_0)/\tau_R},
\end{equation}
with the initial sigma condensate $\sigma(\epsilon,t_0)$ at $t_0$.

\begin{figure}[htbp]
    \centering
        \begin{overpic}[width=0.4\textwidth]{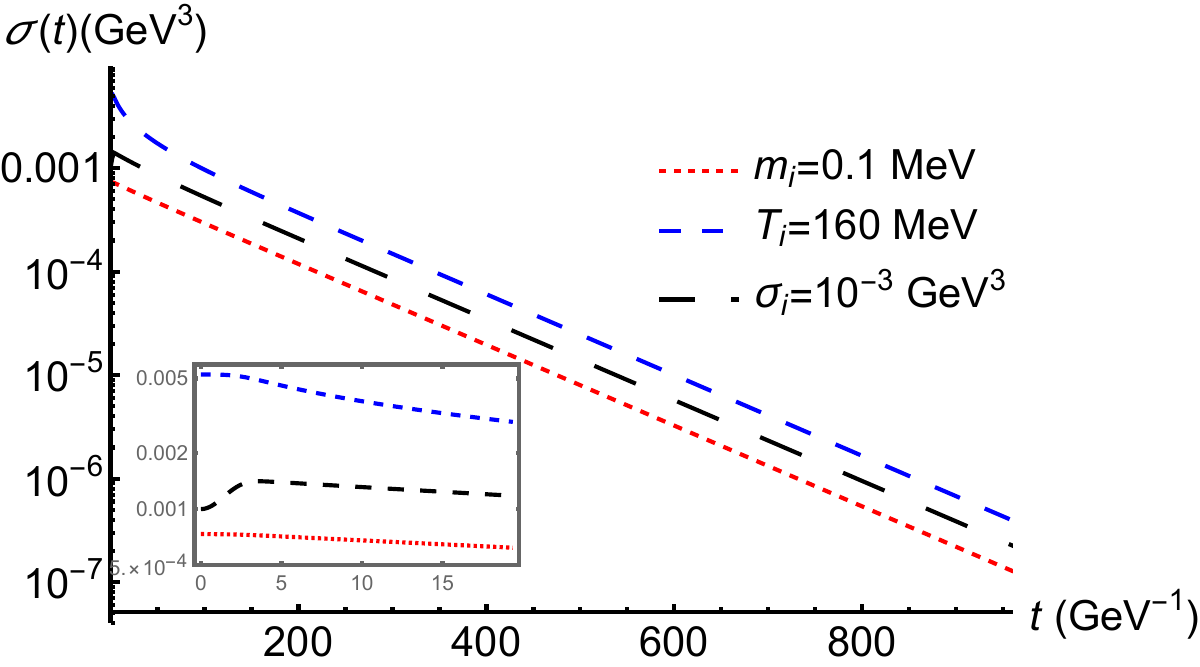}
        \put(85,50){\bf{(a)}}
    \end{overpic}
    \hfill
    \begin{overpic}[width=0.4\textwidth]{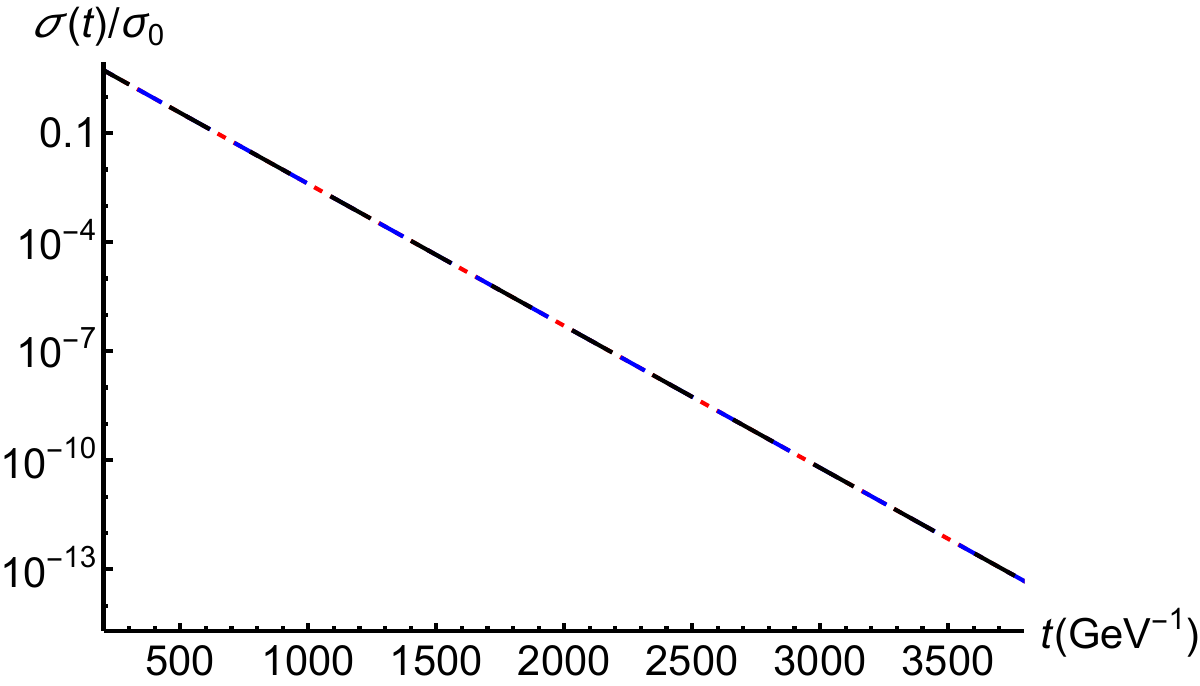}
        \put(85,50){\bf{(b)}}
    \end{overpic}
    \hfill
    \caption{\label{fig:relaxationi} Relaxation of the sigma condensate $\sigma(t)$ with different initial conditions and final temperature $T_f=164$ MeV. \textbf{(a)} Time dependence of sigma condensate $\sigma(t)$. The small inserted figure is a partial enlarged view at the very beginning time region. \textbf{(b)} $\sigma(t)$ rescaled by $\sigma_0=\sigma(200)$}
\end{figure}

\begin{figure}[htbp]
    \centering
    \begin{overpic}[width=0.45\textwidth]{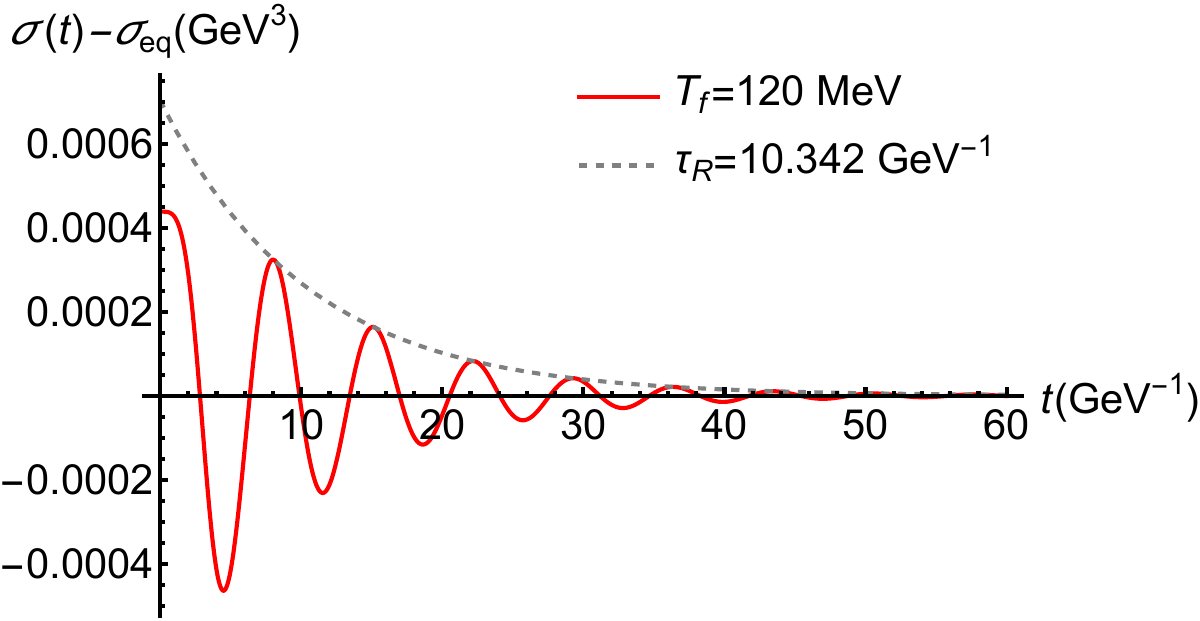}
        \put(85,50){\bf{(a)}}
    \end{overpic}

     \begin{overpic}[width=0.45\textwidth]{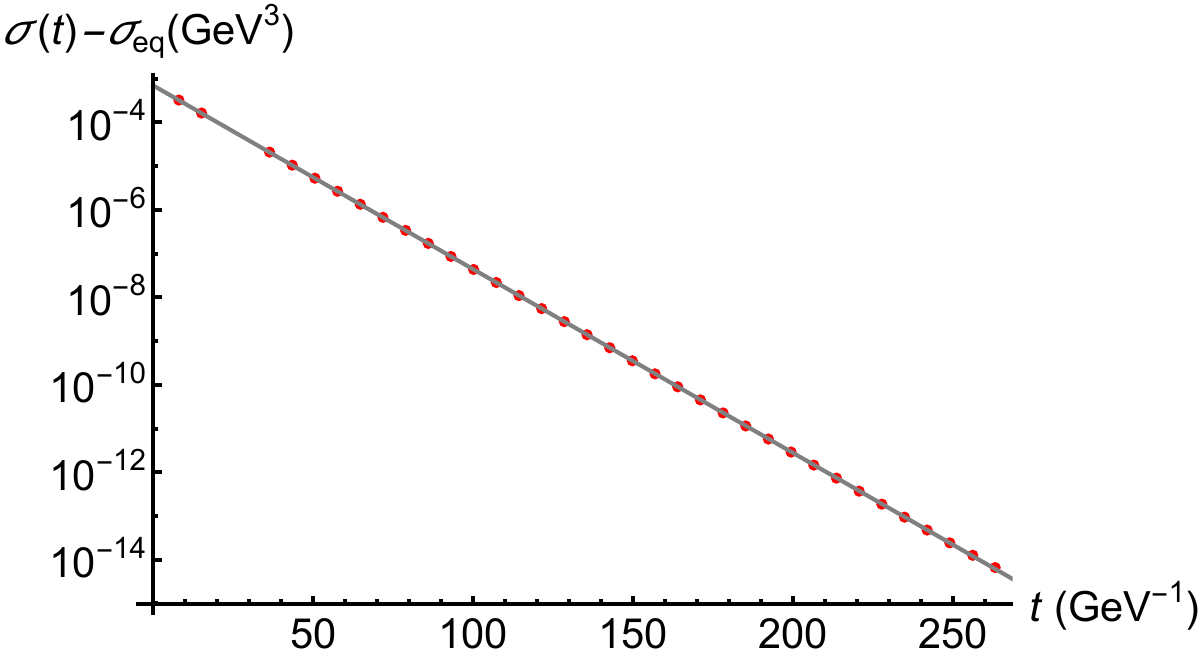}
        \put(85,50){\bf{(b)}}
    \end{overpic}
    \caption{\label{fig:relaxationlowT} \textbf{(a)} Relaxation process of sigma condensate, with initial state at $T_i=110$ MeV and final equilibrium temperature at $T_f=120$ MeV. \textbf{(b)} Envelope of the sigma condensate is fitted with the local maximum values of the oscillation attenuation curve of sigma condensate and the relaxation time $\tau_R=10.342\  {\rm{GeV^{-1}}}$. }
\end{figure}

Generally, to realize the relaxation process described in Eq.~\eqref{evolvingEOMs},  the initial state should slightly deviate from the equilibrium state. We choose a temperature higher than the critical temperature as a final temperature, $T_f=164$ MeV~\footnote{This temperature insure that the final equilibrium sigma condensate $\sigma_{\rm{eq}}=0$}. Generally, the relaxation time $\tau_R$ should not dependent on the details of the initial sate or the way 
by which the initial sate is prepared.  The only physical constrain is the sigma condensate of the initial state should be less than the saturation sigma condensate $\sigma_{sat}$. Here, we will try to verify this numerically. We give three different initial states as examples: one has a tiny quark mass, $m_i=0.1$ MeV; the other one has a temperature slightly below the critical temperature, $T_i=160$ MeV; the last one is given a small quantity of sigma condensate, $\sigma_i=10^{-3}\ {\rm{GeV^3}}$. We numerically calculate the $\chi(\epsilon,t)$ through Eq.~\eqref{eom-chi} with these three different initial states. As presented in Fig.~\ref{fig:relaxationi}(a), after a microscopic timescale ($\sim 2/(\pi T_f)$), the system crossover to the linear relaxation regime, in which sigma condensates behave as exponentially decay (linear
decreasing in the semi-log plot). Then we choose the initial time of the relaxation at $t_0=200\ {\rm{GeV^{-1}}}$ and rescale the time dependence $\sigma(t)$ as shown in Fig.~\ref{fig:relaxationi}(b). The numerical data are well fitted with Eq.~\eqref{evolvingEOMs2} which give relaxation time $\tau_R=111.207, 111.186, 111.202\ {\rm{GeV^{-1}}}$, respectively. Within allowable errors, these three different relaxation processes share the same relaxation time which is determined by the same final state.

When the final state is too far from the critical temperature, the relaxation process behaves differently. For example, we set the system initially at a equilibrium state with $T_i=110$ MeV, then sudden quench to the final temperature $T_f=120$ MeV. The evolution curves of sigma condensate are shown in Fig.~\ref{fig:relaxationlowT}. Instead of pure relaxation, the evolution of $\sigma(t)$ moves in an oscillation damping mode. However, the damping rate of the sigma condensate is determined by the relaxation time. We fitted the envelope curve with Eq.~\eqref{evolvingEOMs2} and get $\tau_R=10.342\ {\rm{GeV^{-1}}}$, as shown in Fig.~\ref{fig:relaxationlowT}(b).

Therefore, we numerically verified that the relaxation is described by the linear response function eqs.~\eqref{evolvingEOMs}, \eqref{evolvingEOMs2}, and the relaxation time is related to the distance deviated from the critical temperature $\epsilon=T_f-T_c$ in the soft-wall model\footnote{The relaxation time also is a function of quark mass and other external parameters}.

\subsubsection{Critical slowing down}\label{ctsd}
In the thermodynamic limit, it is well known that the critical slowing down arises at the critical point, since the correlation length is getting divergent, as well as the relaxation time. However, this nonequilibrium phenomenon in the holographic QCD still lacks sufficient investigations. In this section, we will verify the critical slowing down phenomenon with the dynamical evolution of the order parameter $\sigma(t)$ in the soft-wall model.

In the critical region (near  the critical point), the correlation length $\xi$ should satisfy the scaling hypothesis~\cite{zinn2021quantum}:
\begin{equation}\label{scalingxi}
    \xi(\epsilon, m_q, t)= b  \xi(\epsilon b^{1/\nu}, m_q b^{\beta \delta/\nu}, t b^{-z}),
\end{equation}
with an additional length rescaling factor $b$ and static critical exponents $\beta$, $\nu$ and $\delta$. From Eq.~\eqref{scalingxi}, it is implicated that the sigma condensate satisfies
\begin{eqnarray}\label{eq:sigmascaling}
\sigma(\epsilon,m_q,t)=b^{-\beta/\nu}\sigma(\epsilon b^{1/\nu},m_q b^{\beta\delta/\nu},t b^{-z}),
\end{eqnarray}
and the relaxation time $\tau_R$ behaves as
\begin{eqnarray}
\tau_R(\epsilon,m_q,t)=b^{z}\tau_R(\epsilon b^{1/\nu},m_q b^{\beta\delta/\nu},t b^{-z}).
\end{eqnarray}
After choosing particular scaling parameters, $m_q$ and $\epsilon$, the corresponding scaling forms can be derived and the leading orders of the scaling forms, respectively, behave as
\begin{subequations}\label{eqpower}
\begin{eqnarray}
\xi &\sim& m_q^{-\nu/\beta \delta},\quad 
\sigma \sim m_q^{1/\delta},\quad \quad
\tau_R \sim m_q^{-\nu z/\beta \delta},\\
\xi &\sim& \epsilon^{-\nu},\quad \quad \quad 
\sigma \sim \epsilon^{\beta},\quad \quad\quad
\tau_R \sim \epsilon^{-\nu z}.
\end{eqnarray}
\end{subequations}
Those static critical exponents, $\beta$, $\delta$ and $\nu$, have been obtained numerically and analytically in the previous Refs.\cite{Chen:2018msc,Cao:2021tcr} with the soft-wall model. However, to describe the evolution of thermalization, an additional dynamic critical exponent $z$ is required. From Eq.~\eqref{eq:sigmascaling}, one can obtain the scaling form in terms of $t$ as
\begin{equation}
    \sigma(t)=t^{-\beta/\nu z}f_{th}(\epsilon t^{1/\nu z}, m_q t^{\beta\delta/\nu z}),
\end{equation}
with the scaling function $f_{th}$. So that the sigma condensate decays as a power law of the form
\begin{equation}\label{criticalslowingdown}
    \sigma(t)=t^{-\beta/\nu z}f_{th}(0,0)\propto t^{-\beta/\nu z},
\end{equation}
at the critical point, $\epsilon=0$ and $m_q=0$. It means that infinite time is needed to recovery equilibrium state $\sigma=0$ at the critical point, which is the famous critical slowing down phenomena.


\begin{figure}[htbp]
    \centering
        \begin{overpic}[width=0.45\textwidth]{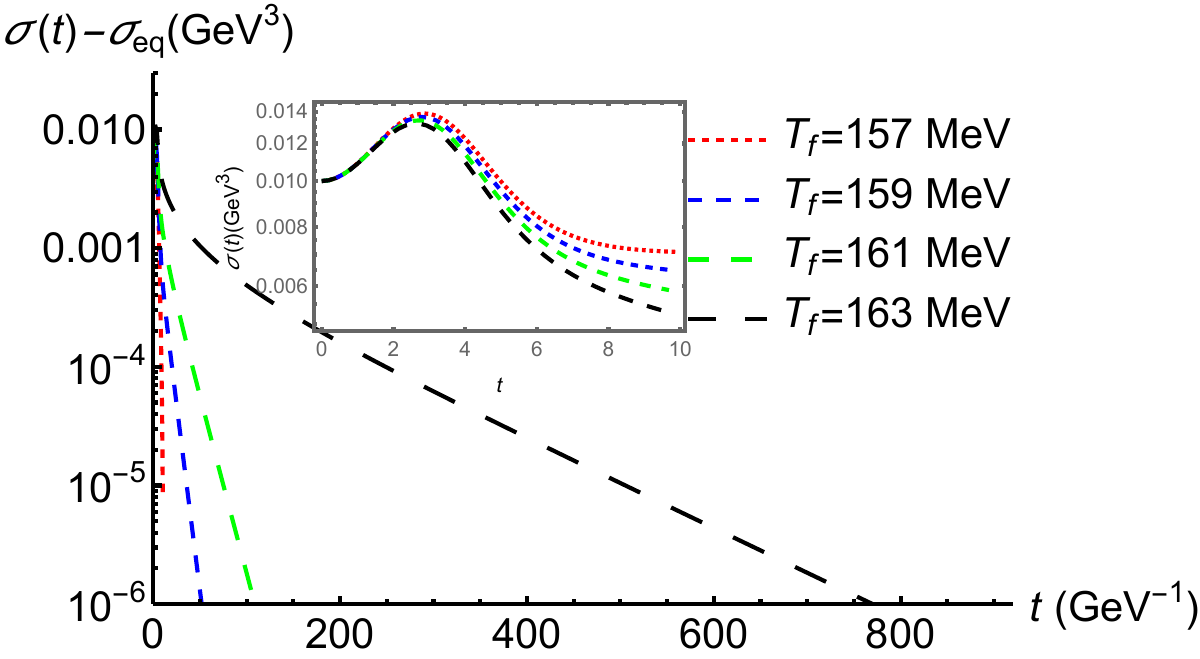}
        \put(80,50){\bf{(a)}}
    \end{overpic}
    \begin{overpic}[width=0.43\textwidth]{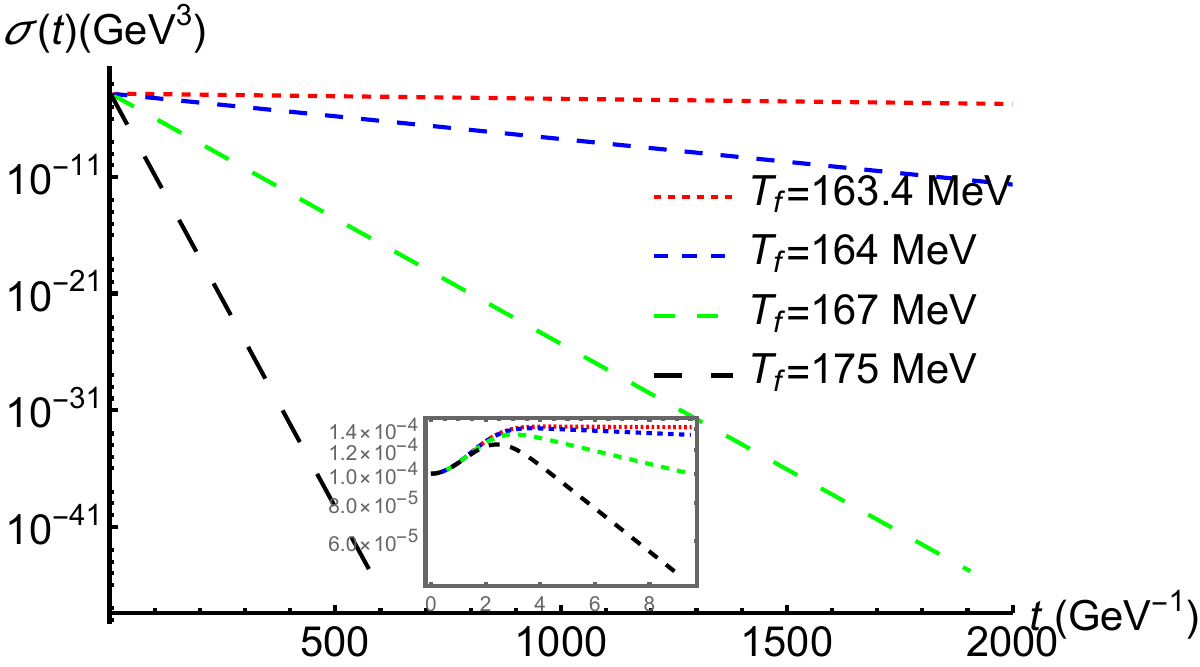}
        \put(81,50){\bf{(b)}}
    \end{overpic}
    \caption{\label{fig:sigmavt} Evolution of sigma condensate $\sigma(t)$ with different final temperatures $T_f$. a) Time dependence of $\sigma(t)-\sigma_{\rm{eq}}$ with the same initial state $\sigma_0=10^{-2}      \ {\rm{GeV}}^3$ quenched to different final temperatures $T_f$ in the ordered phase. b) Time dependence of $\sigma(t)$ with the same initial state $\sigma_0=10^{-4} {\rm{GeV}}^3$ quenched to different final temperatures $T_f$ in the disordered phase.}
\end{figure}

We adopt four cases, they begin with the same initial sigma condensate. Since it is verified in the last subsection, the initial values have no effects on the late lime relaxation process, we only take a specific case of $\sigma_0=10^{-2}\ {\rm{GeV^3}}$ as an example. Then, one can quench the initial state to  different final temperatures, $T_f=157,\ 159, \ 161$ and $163$ MeV, as shown in Fig.~\ref{fig:sigmavt}(a). From Eq.~\eqref{evolvingEOMs2}, one has the slop of the semi-logarithmic curve corresponding to the inverse of the relaxation time. By fitting the data, we obtain $\tau_R=3.960, 6.929, 14.541, 113.964 \ {\rm{GeV^{-1}}}$ for $T_f=157, 159, 161, 163$ MeV, respectively. Varying the final temperature $T_f$ in the condensed phase, the relaxation time increases with $T_f$. On the other hand, the relaxation of sigma condensate in the chiral symmetry restored phase is also shown in Fig.~\ref{fig:sigmavt}(b). We extract the relaxation time $\tau_R=980.566, 111.205, 20.177, 6.092 \ {\rm{GeV^{-1}}}$ for $T_f=163.4, 164, 167, 175$ MeV, respectively. The relaxation time decreases with the increasing temperature for $T_f>T_c$. Therefore, the relaxation time diverges either approaching to or receding from $T_c$.

\begin{figure}[htbp]
    \centering
    \begin{overpic}[width=0.45\textwidth]{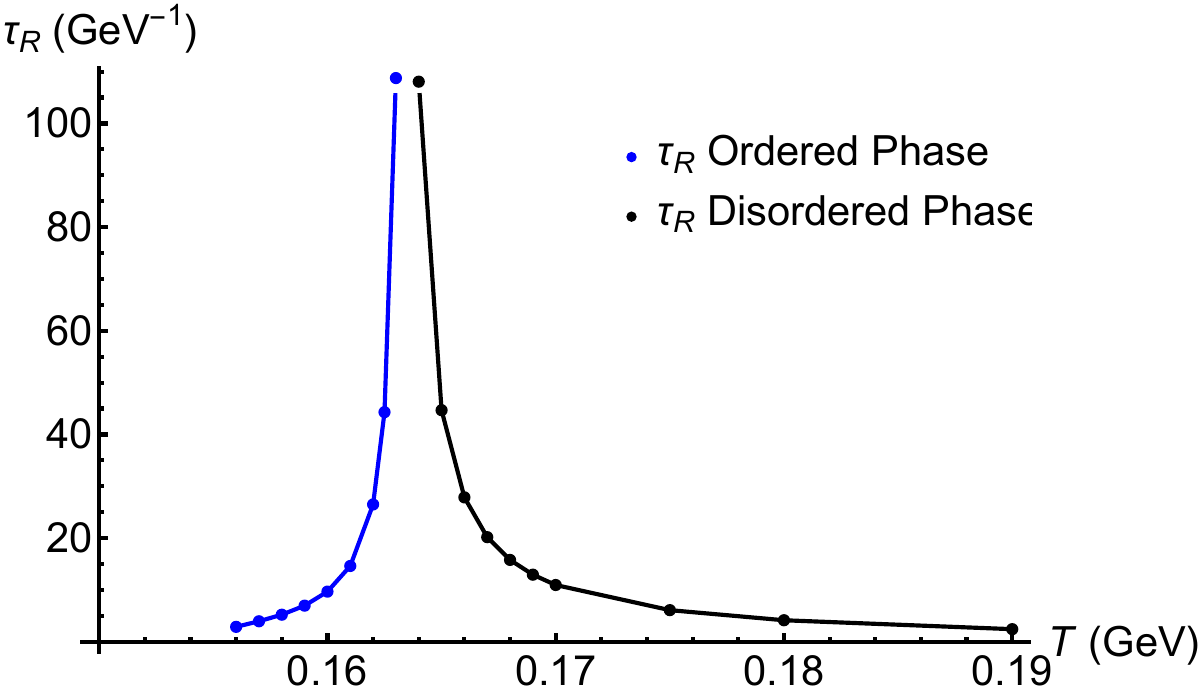}
        \put(85,50){\bf{(a)}}
    \end{overpic}
    \begin{overpic}[width=0.45\textwidth]{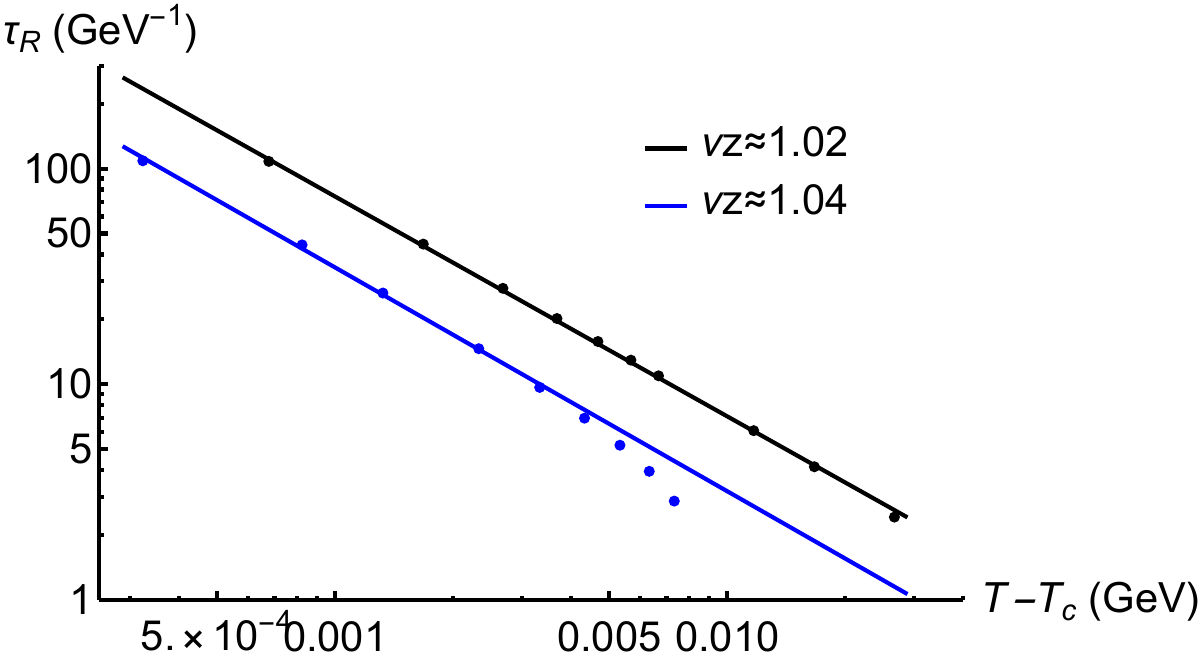}
        \put(85,50){\bf{(b)}}
    \end{overpic}
    \caption{\label{fig:relaxationtime} \textbf{(a)} Relaxation time constant $\tau_R$ as a function of temperature.\textbf{(b)} Fitting the relaxation time constants with Eq.~\eqref{eqpower}(b).}
\end{figure}

Furthermore, we show extracted values of relaxation time at different temperatures in Fig.~\ref{fig:relaxationtime}. The result shown that the relaxation time diverges at the critical temperature. When the system is a certain distance deviation from the critical point, the relaxation of the sigma condensation decreases exponentially, satisfying Eq.~\eqref{evolvingEOMs2} with a particular relaxation time, $\tau_R$. As shown in Fig.~\ref{fig:relaxationtime}b, when $\epsilon$ approaches zero, the relaxation time
behaviors as the power-law divergence form in Eq.~\eqref{eqpower}b. We fitted out the combination exponents, $-\nu z=-1.04$ and $-1.02$ for ordered phase and disordered phase, respectively. Since $\nu=1/2$, we have $z\approx 2.08$ and $2.04$.
\begin{figure}[htbp]
    \centering
        \begin{overpic}[width=0.45\textwidth]{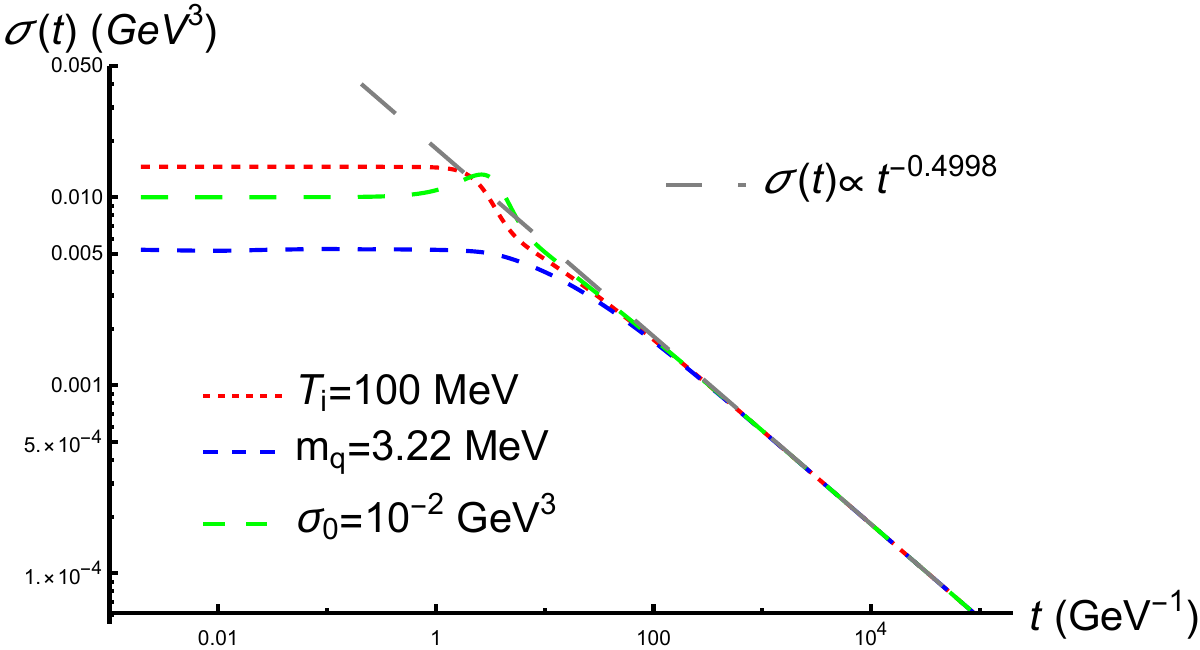}
        \put(85,50){\bf{(a)}}
    \end{overpic}
        \centering
        \begin{overpic}[width=0.45\textwidth]{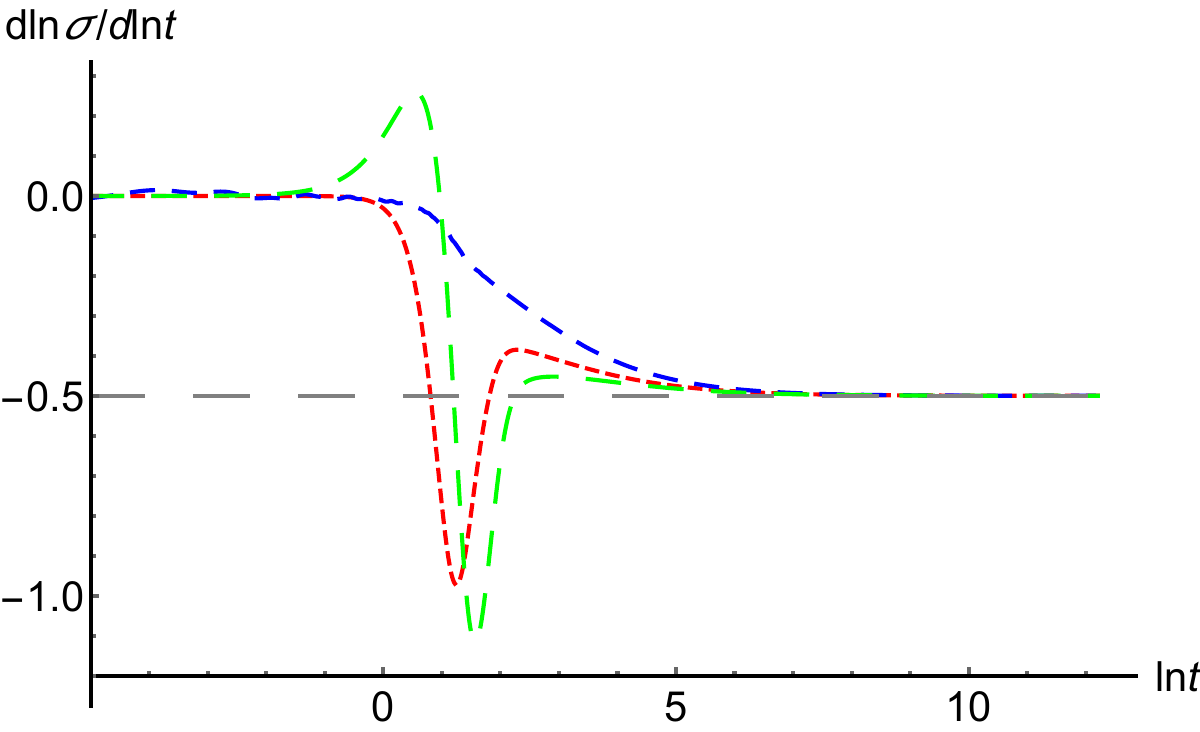}
        \put(85,53){\bf{(b)}}
    \end{overpic}
    \caption{\label{fig:critialrelaxation} \textbf{(a)} Time dependent relaxation of $\sigma$ quenched to the critical temperature with different initial states. \textbf{(b)} Time dependent evolution of $d\ln{t}/ dt$}
\end{figure}
Except the correlation time, we can study the critical slowing down phenomena directly from the order parameter. As shown in Fig.~\ref{fig:critialrelaxation}(a), As examples, we have three different evolution curves of $\sigma(t)$ with different initial states quenched to the critical point. These different initial states are initial temperature $T_i=100$ MeV, initial quark mass $m_i=3.22$ MeV, and initial sigma condensate $\sigma_i=10^{-2}\  {\rm{GeV^3}}$. After a very initial time stage, which mainly depends on the initial configurations and the microscopic details, sigma condensate relaxes to equilibrium state in an extremely long time, as shown in Fig.~\ref{fig:critialrelaxation}(b) \footnote{During the review process, it is interesting to see that the recent study in holographic superfluid shows a similar behavior \cite{Flory:2022uzp}, though the symmetries considered in the two systems are different. }. The slope of $\sigma (t)$ in the log-log plot is about a constant value $-0.4998$ at the long time relaxation stage. From Eq.~\eqref{criticalslowingdown}, it yields $-\beta/\nu z\approx-0.4998$. Thus, the dynamic exponent $z\approx 2.0008$, which is consistent with the result obtained from the scaling of the relaxation time. According to the classification in Ref.~\cite{hohenberg1977theory}, the soft-wall model belongs to Model A. ~\footnote{Since the critical slowing down, it is available to determine the critical temperature with their scaling behaviors of the relaxation time or the sigma condensate. However, the critical temperature with extremely high accuracy is obtained through the static method proposed in Ref.~\cite{Chen:2018msc, Cao:2020ryx}.}

\subsection{The relation between the thermalization and QNM}
At finite temperature, the Lorentz symmetry is broken, and the real part of the dispersion relation would behave as
\begin{equation}
\{{\rm Re}[\omega(\mathbf{p})]\}^2=u^2_{\pi} (\mathbf{p}^2+m_{\rm{scr}}^2).
\end{equation}
$u_{\pi}$ is the pion velocity; $m_{\rm{scr}}$ is the screening mass which satisfy $\mathbf{p}^2=-m_{\rm{scr}}^2$ at $\omega=0$; and $m_{\rm{pole}}={\rm Re}[\omega(0)]=u_{\pi} m_{\rm{scr}}$ is the pole mass at $\mathbf{p}=0$ ~\cite{Son:2001ff,Son:2002ci}. 

The quasinormal mode is the oscillation mode of the perturbation of the background. The QNM frequency $\omega_0$ corresponds to the pole of two-point Green's function at $\mathbf{p}=0$~\cite{Son:2002sd}.  Under the framework of holographic duality, the real and imaginary part of $\omega_0$ correspond to the pole mass $m_{\rm{pole}}$ and the thermal width $\Gamma$ ~\cite{Miranda:2009uw}. We have already verified these relationships ($m_{\rm{pole}}={\rm{Re}}[\omega_0]$ and $\Gamma/2=-{\rm Im}[\omega_0]$) in the soft-wall AdS/QCD model in our previous work~\cite{Cao:2021tcr}. In that work, we also talk about the screening mass $m_{\rm{scr}}$ (inverse of the correlation length $\xi^{-1}$). A particular momentum satisfying $\mathbf{p}_0^2+m_{\rm{scr}}^2=0$, corresponds to the pole of the two-point retarded Green's function at $\omega=0$.

In this section, we will study the relationships among the relaxation time, the correlation length, the screening mass, and the QNM frequency. For completeness, we briefly review the derivations of the retarded Green's function of the scalar mode, more details of the derivation in Ref.~\cite{Son:2002sd}. The perturbation action of the scalar sector is,
\begin{eqnarray}\label{scalar-T}
S_{\sigma}&=&\frac{1}{2}\int d x^5 \sqrt{g}e^{{-\Phi}}\bigg [ g^{\mu\nu}\partial_{\mu}S\partial_{\nu} S+g^{rr}(\partial_r S)^2-\nonumber\\
& &m_5^2 S^2-\frac{3\lambda}{2}\chi^2 S^2\bigg ].
\end{eqnarray}
Then, we can derive the EOM for the scalar meson $S$ as
\begin{eqnarray}\label{EOM:scalar}
    S''+\left(3 A'+\frac{f'}{f}-\Phi'\right)S' +& &\nonumber \\
    \left(\frac{\omega^2}{f^2}-\frac{p^2}{f}-\frac{2 m_5^2+3 \lambda \chi ^2}{2 f}A'^2 \right)S&=&0,
\end{eqnarray}
which is transformed into the momentum space $(\omega, \mathbf{p})$. For simplicity, we let $\mathbf{p}$ along a particular $x_1$-direction $\mathbf{p}=(p,0,0)$.

Near the boundary at $r=0$, one can obtain the boundary asymptotic expansion as
\begin{equation}
    S(r)=s_1 r+s_3 r^3+\cdots,
\end{equation}
with two integral constants $s_1$ and $s_3$. According to the holographic dictionary, one has $s_1$ corresponding to the extra source $J_s$. The incoming wave condition at the horizon $r=r_h$ is
\begin{equation}
    S(r)\sim (r-r_h)^{-i\omega t/(4\pi T)}.
\end{equation}
Combining these conditions, we numerically solve the EOM Eq.~\eqref{EOM:scalar} though the so-called ``shooting method''~\cite{boyd2001chebyshev,Cao:2020ryx}. 

\begin{figure}[htbp]
    \centering
     \begin{overpic}[width=0.45\textwidth]{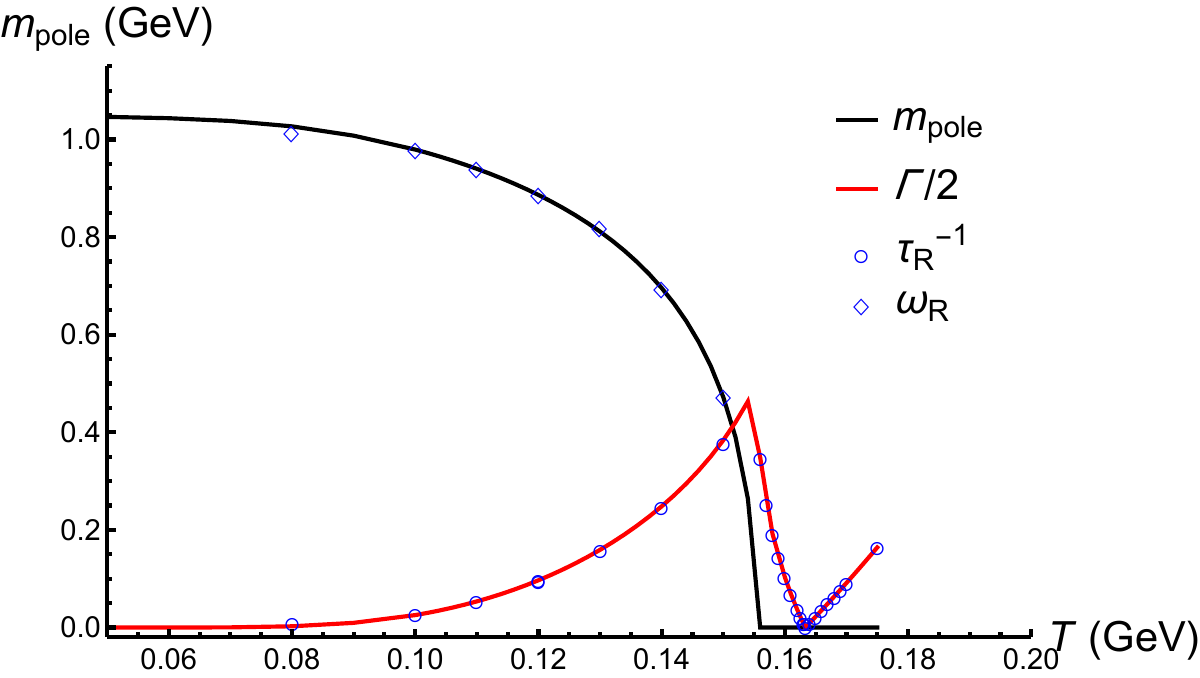}
        \put(85,50){\bf{(a)}}
    \end{overpic}
    \begin{overpic}[width=0.45\textwidth]{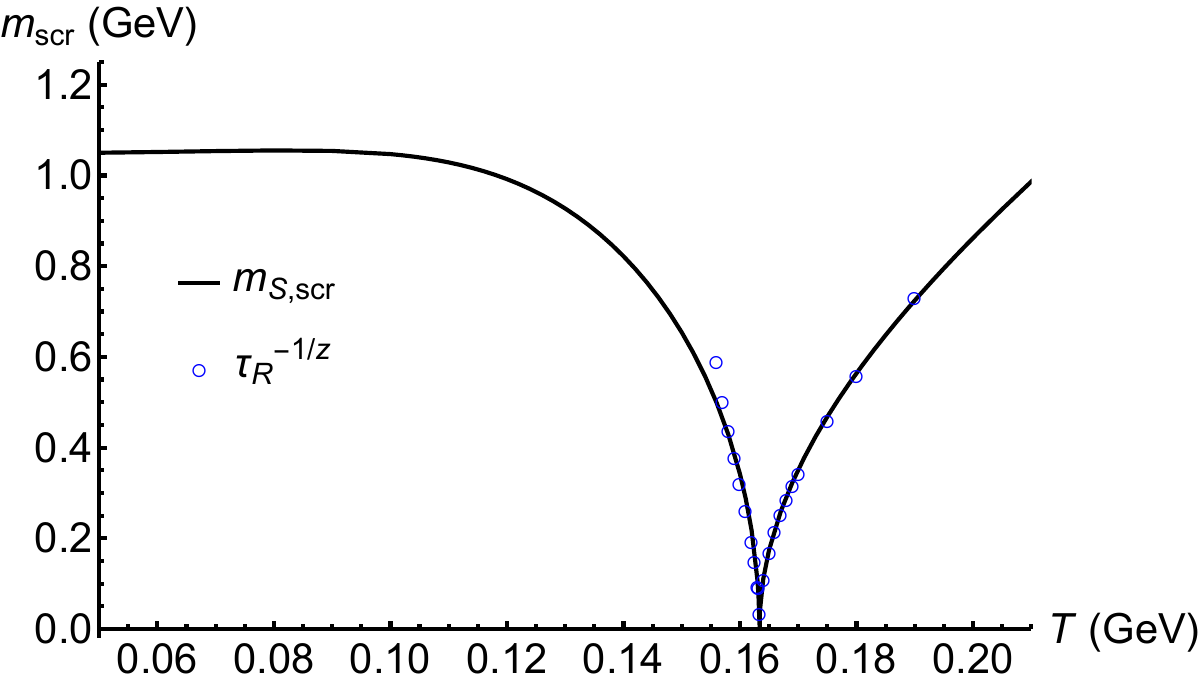}
        \put(85,49){\bf{(b)}}
    \end{overpic}
    \caption{\label{fig:QNMcase2}  \textbf{(a)} Comparison between the thermal width and the relaxation time, the pole mass and the relaxation frequency. It is shown that $\Gamma_S/2\approx \tau_R^{-1}$ and $m_{\rm{pole}}\approx \omega_R$. \textbf{(b)} Comparison between the screening mass and the relaxation time. In the critical region, it is shown that $m_{scr}=\xi^{-1}\approx k\tau_R^{-1/z}$ with a scalar factor $k\approx 2.47444 \ T_c^{1/2}$ in the ordered phase and $k\approx 2.82086 \ T_c^{1/2}$ in the disordered phase.}
\end{figure}

Following the prescription in Ref.~\cite{Son:2002sd}, One has the retarded Green's function of $S$ proportional to the ratio of $s_1$ and $s_3$,
\begin{equation}\label{eq:green'sfunction}
    G_s(\omega, p)\sim \frac{s_3(\omega, p)}{s_1(\omega, p)}.
\end{equation}
From Eq.~\eqref{eq:green'sfunction}, we know that the pole of the two point retarded Green's function is equivalent to $s_1(\omega, p)=0$. To obtain the pole mass $m_{\rm{pole}}$ and the thermal width $\Gamma$, we need solve $s_1(\omega, 0)=0$ and label the solution of the frequency as $\omega=\omega_0$. As to the screening mass $m_{\rm{scr}}$, we should solve $s_1(0,p)=0$ and label the solution of the momentum as $p=p_0$.

We show the numerical results of pole mass \footnote{We remind that, at low temperatures, the finite pole mass induces oscillates along with the relaxation evolution, as the case shown in Fig.~\ref{fig:relaxationlowT}.}, thermal width, and the screening mass in Fig.~\ref{fig:QNMcase2}(a) and the screening mass and the relaxation time in Fig.~\ref{fig:QNMcase2}(b). The thermal width curve collapses with the curve of the inverse of the relaxation time. It means that the inverse of the thermal width can be identified as the relaxation time, i.e., $\Gamma/2\approx 1/\tau_R$.
 As shown in Fig.~\ref{fig:relaxationi} and \ref{fig:relaxationlowT}, the value of sigma condensate is varying with time in the relaxation process. More importantly, the absolute value of the sigma condensate departing from the equilibrium value is very small and approaching to zero. It means that the whole relaxation process can be regarded as a slight perturbation on the equilibrium state. From the aspect of the scalar meson, it is enough to only consider up to the quadratic terms in the perturbation action ~\eqref{scalar-T}. That would be the reason why the relaxation mode is well consistent with scalar meson mode.
As shown in Fig.~\ref{fig:QNMcase2}(b), based on Eq.~\eqref{eqpower}(b), it is verified that the scaling mass and the relaxation time satisfy the relation $m_{\rm{scr}}=\xi^{-1}\approx k\tau_R^{-1/2}$ in the critical region with $k$ a fitting parameter. In our numerical results, the fitting results are $k\approx 2.47444 \ T_c^{1/2}$ in ordered phase and $k\approx 2.82086 \ T_c^{1/2}$ in disordered phase.

\section{Prethermalization}\label{sec:prethermalization}
Governed by the time scales, the nonequilibrium process can be naturally dived into three stages, including the microscopic timescale dominated prescaling stage at the very beginning period, the prethermalization in the intermediate time, and the long time thermalization. For the third stage, it has been studied in the last section. Here, we will focus on the short-time dynamics arising in the intermediate time.

\subsection{Quench protocols}
\begin{figure}[tbp]
  \centering
	\begin{overpic}[width=0.45\textwidth]{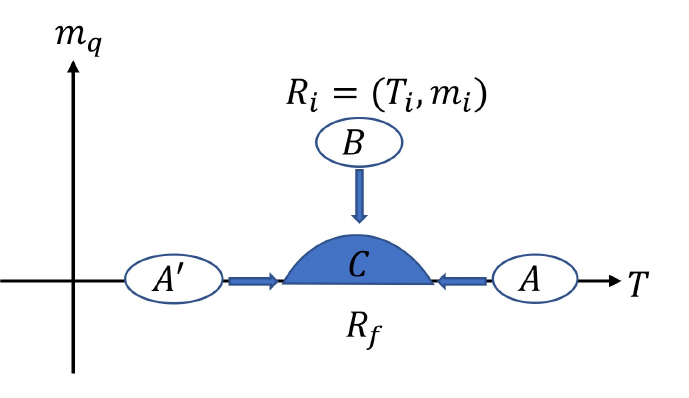}
               \put(85,50){\bf{(a)}}
         \end{overpic}
	\begin{overpic}[width = 0.45\textwidth]{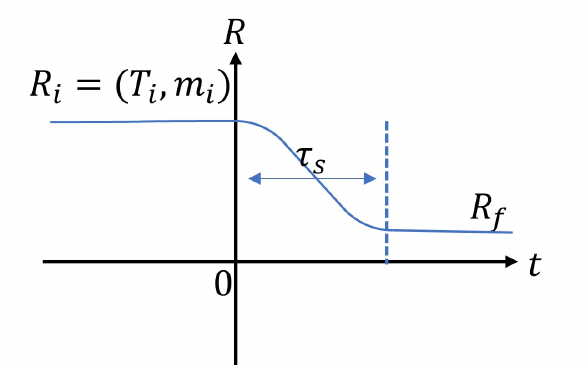}
		\put(80,50){\bf{(b)}}
	\end{overpic}
\caption{\textbf{(a)} Schematic diagram of quench protocols. \textbf{(b)} Time dependence of the external parameters $R(t)$.}
\label{sketch}
\end{figure}

To clearly reveal the short-time dynamics, we have a sketch for the quenching protocols. In the nonequilibrium evolution, the external parameters change with time,
\begin{equation}
    R(t)\equiv (T(t),m(t)).
\end{equation}
The initial state of the system is elaborately prepared, $R(t=0)\equiv R_i=(T_i,m_i)$. Eventually, the system is quenched to the critical region, $R_f\equiv R(t\rightarrow +\infty)=(T_f, m_q)$. As shown in Fig.~\ref{sketch}(a), three different initial states will be considered. Those are extremely high temperate with a finite sigma condensate ($A\rightarrow C$), finite quark mass ($B\rightarrow C$) and the equilibrium state in the ordered phase ($A'\rightarrow C$). The changing of the external parameters is shown in Fig.~\ref{sketch}(b). To quench from $R_i$ at $t=0$ to $R_f$ is within a finite timescale $\tau_Q$. In this work, we only consider the sudden quench case, i.e., $\tau_Q\rightarrow 0$, so that
\begin{equation}
    R(t)=R_i+\theta(t)(R_f-R_i),
\end{equation}
with $\theta(t)$ the step function.

\subsection{Short time scaling}
In section~\ref{sec:thermalization}, we have studied the properties in the limit of thermalization. In the asymptotic long-time stage, all the initial state information has been washed out. It is shown that the system reaches the equilibrium thermal state, which only depends on the final external parameters $R_f$. However, in the prethermalization stage, the evolution is expected to depend on the initial parameters, $R_i=(T_i,m_i)$. Be inspired by the short-time scaling in the condensed matter model~\cite{janssen1989new,Schoeller:2009ape,Berges:2000ew,Gagel:2015opa}, we have a suggested scaling hypothesis for the sigma condensate,
\begin{equation}\label{STscaling}
    \sigma(R_i,\epsilon,m_q,t)=b^{-\beta/\nu}\sigma(R_i(b),\epsilon b^{1/\nu},m_q b^{\beta\delta/\nu},tb^{-z}),
\end{equation}
with a scaling length parameter $b$. $R_i(b)$ is the initial parameters rescaled by $b$. As a result of the memory of the initial state, the scaling is proposed to employ a new exponent $x$.  And Eq.~\eqref{STscaling} becomes
\begin{widetext}
\begin{equation}\label{scalingepsilon}
    \sigma(\epsilon_i,m_i,\epsilon,m_q,t)=b^{-\beta/\nu}\sigma (\epsilon_i b^{x/\nu}, m_i b^{x\beta\delta/\nu},\epsilon b^{1/\nu}, m_q b^{\beta\delta/\nu},tb^{-z}),
\end{equation}
\end{widetext}
Near the critical point, the equilibrium sigma condensate relates to the distance to critical point with $\sigma\sim \epsilon^{\beta}$, so that one obtains
\begin{widetext}
\begin{eqnarray}\label{scalingsigmai}
     \sigma(\sigma_i,m_i,\epsilon,m_q,t)=b^{-\beta/\nu}\sigma (\sigma_i b^{x\beta/\nu}, m_i b^{x\beta\delta/\nu},\epsilon b^{1/\nu}, m_q b^{\beta\delta/\nu},tb^{-z}).
\end{eqnarray}
\end{widetext}

\subsubsection{Quench from the disordered phase}
\begin{figure*}[htb]
    \centering
    \begin{overpic}[width=0.31\textwidth]{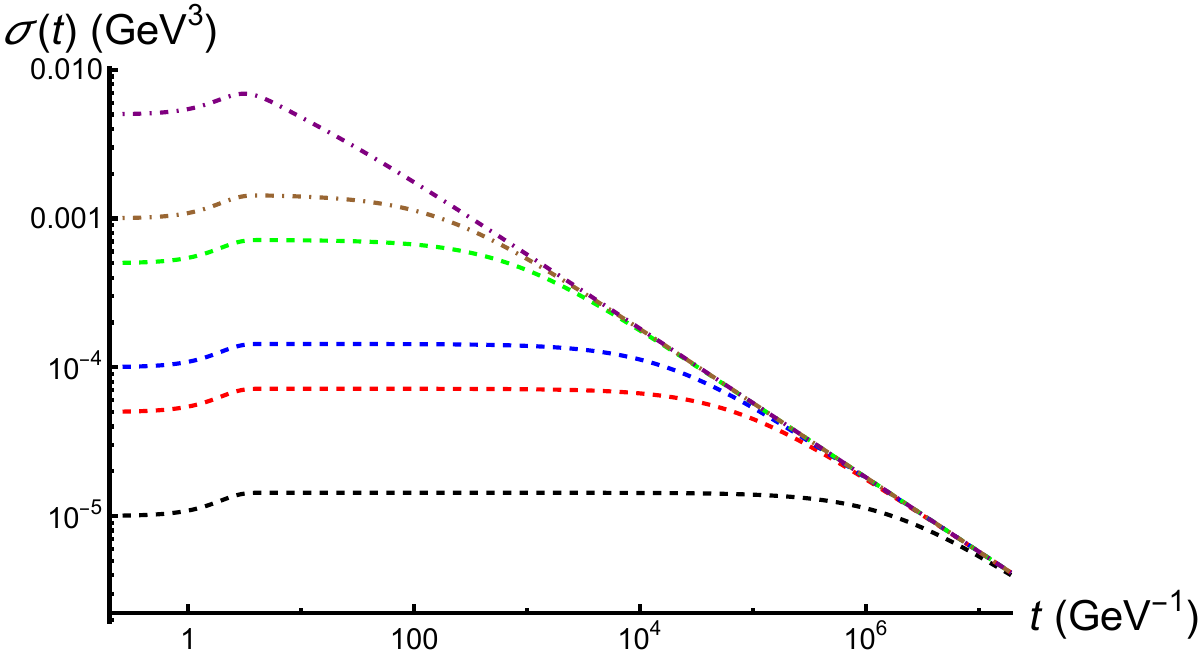}
    \put(90,50){\bf{(a)}}
    \end{overpic}
     \begin{overpic}[width=0.31\textwidth]{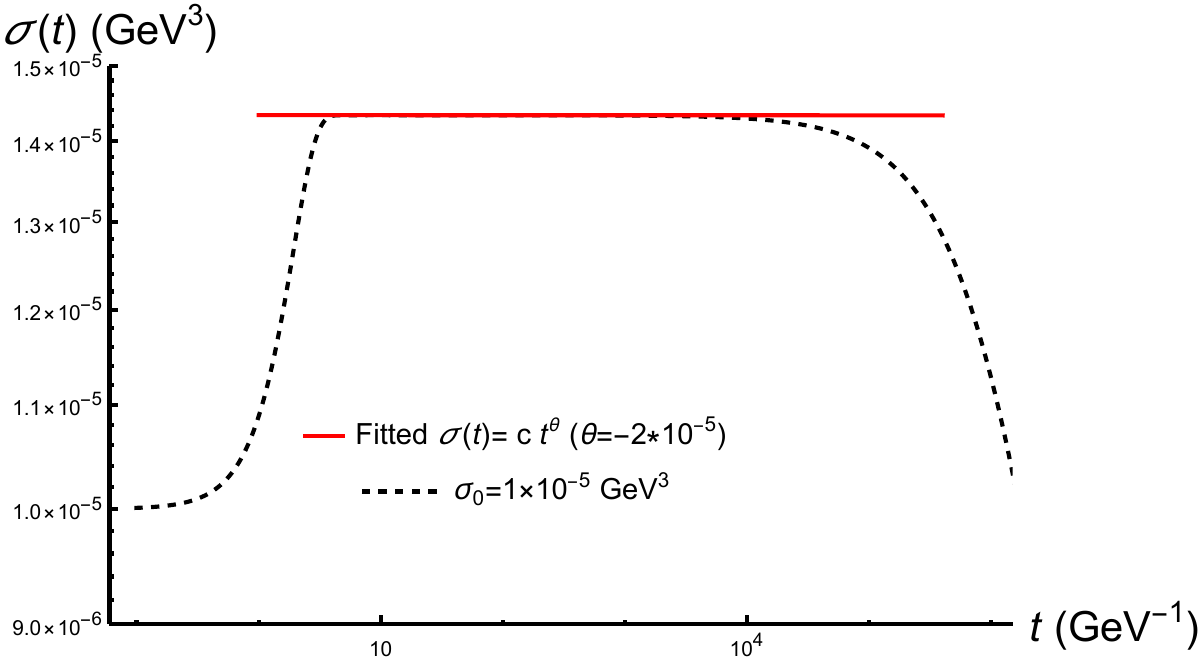}
    \put(90,50){\bf{(b)}}
    \end{overpic}
    \begin{overpic}[width=0.31\textwidth]{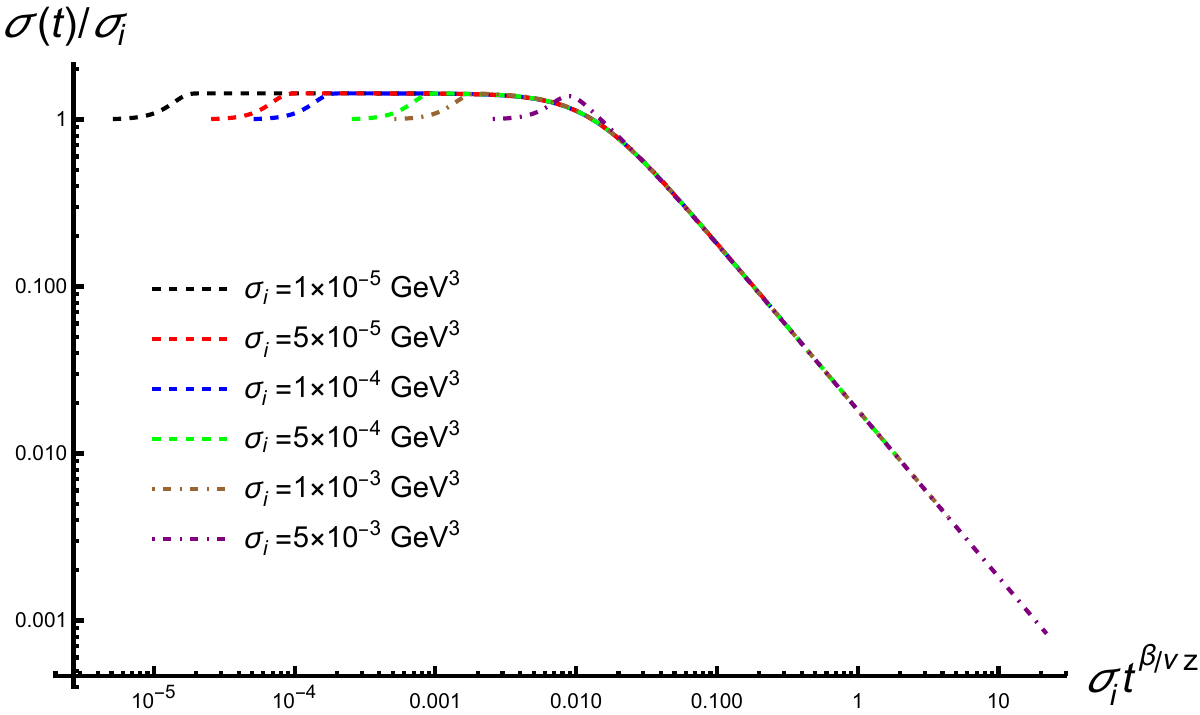}
        \put(90,50){\bf{(c)}}
    \end{overpic}
        \caption{\label{fig:8} \textbf{(a)} Evolution of the sigma condensate with different initial sigma condensate values sudden quenched to the critical point; \textbf{(b)} Fitting the intermediate time with Eq.~\eqref{scalingst}, one get the short-time dynamic exponent $\theta=-2*10^{-5}$; \textbf{(c)} Scaling the evolution curves of sigma condensate in (a) based on Eq.~\eqref{scalingsigma}}
\end{figure*}
In the chiral limit, when the temperature is above the critical temperature $T>T_c$, the solution of $\chi(z)$ is exactly equals to $0$. Thus the sigma condensate $\sigma=0$. To realize the nonequilibrium evolution from the disordered phase to the critical region, a finite initial condensate $\sigma_i$ is necessary for the initial state. From the asymptotic solution of $\chi(r)$ in Eq.~\eqref{eq:boundaryofchi}, we know that $m_q$ and $\sigma$ are two integral constants for this solution. Since the chiral limit $m_q=0$, the leading term of this solution becomes $(\sigma/ \gamma) r^3$. In addition, $r_h=1/T$ is very small at high temperature. It means that $r\leq r_h$ is very small. Therefore, when $T_i>T_c$, we can take the asymptotic solution as a approximation of the initial state, 
\begin{equation}
    \chi(r)=\frac{\sigma_i}{\gamma} r^3.
\end{equation}
Firstly, we consider the sudden quench to the critical point, i.e., $\epsilon=T_f-T_c=0$ and $m_q=0$. To derive the specific scaling form in terms of $t$, one can let $tb^{-z}=t_{pre}$ with $t_{pre}$ a microscopic time scale. $t_{pre}$ marks the moment when the universal prethermalization stage begins. Then, from Eq.~\eqref{scalingsigmai}, we have
\begin{equation}
     \sigma(\sigma_i,0,0,0,t)=t^{\frac{-\beta}{\nu z}}\sigma \left[\sigma_i \left({t}/{t_{pre}}\right)^{\frac{x\beta}{\nu z}}, 0,0, 0,t_{pre}\right].
\end{equation}
One can define a new scaling function
\begin{equation}
f_t(\sigma_i t^{x\beta}/{\nu z})\equiv \sigma \left[\sigma_i (t/t_{pre})^{{x\beta}/{\nu z}}, 0,0, 0,t_{pre}\right ].
\end{equation}

Therefore, one can obtain the following scaling form in terms of $t$ as
\begin{equation}\label{scalingintermsoft}
    \sigma(\sigma_i,t)= t^{-\beta/\nu z}f_{t}(\sigma_i t^{x\beta/\nu z}).
\end{equation}
In the long-time region, the system evolves into the thermalization stage and $\sigma_i t^{x\beta/\nu z}\gg 1$, so that $f_{t}=\text{Const}$ and $\sigma(\sigma_i,t)\propto t^{-\beta/\nu z}$, which has been verified in Sec.~\ref{sec:thermalization}. In the short-time region, one has $t\ll t_{th}\propto \sigma_i^{-\nu z/x\beta}$. The time scalar $t_{th}$ marking the system crossover to the thermalization regime. Note that $t$ is often referred to as waiting time. In this period, the magnitude of sigma condensates hasn't varied too much, compared to $\sigma_i$, so that $f_{t}$ is dominated by its linear term. It implicates that
\begin{equation}\label{scalingst}
    \sigma(\sigma_i,t)\propto \sigma_i t^{(x-1)\beta/\nu z}.
\end{equation}
Besides, the general scaling form in terms of $\sigma_i$ is presented by Eq.~\eqref{scalingsigmai} as
\begin{equation}\label{scalingsigma}
    \sigma(\sigma_i,t)=\sigma_i^{1/x} f_{\sigma_i}(t\sigma_i^{z\nu /x\beta}),
\end{equation}
with a scaling function $f_{\sigma_i}$.
In the short-time, one has $f_{\sigma_i}\propto \sigma_i^{(x-1)/x}t^{(x-1)\beta/\nu z}$ to guarantee that $\sigma(\sigma_i, t)$ satisfies Eq.~\eqref{scalingst}. One can define the short time dynamic (dynamic initial-slip) exponent $\theta\equiv (x-1)\beta/\nu z$ to characterizes the universal short-time behavior.

\begin{figure}[phtb]
    \centering
    \begin{overpic}[width=0.42\textwidth]{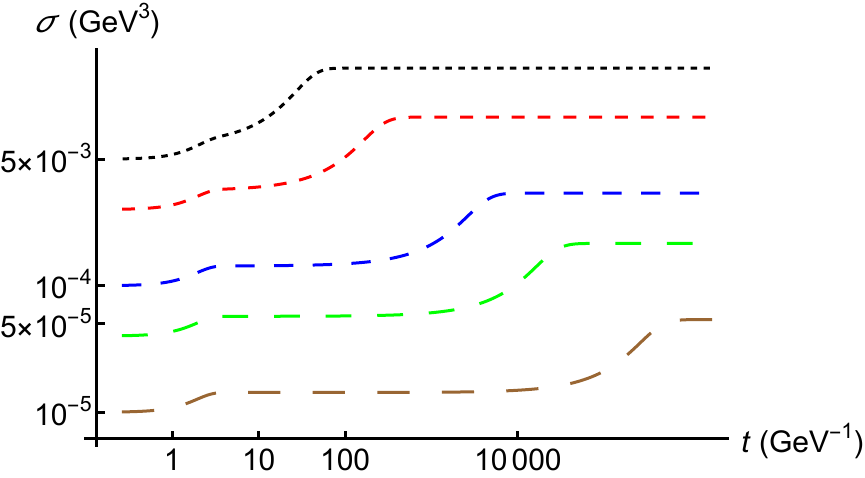}
        \put(85,50){\bf{(a)}}
    \end{overpic}
    \begin{overpic}[width=0.42\textwidth]{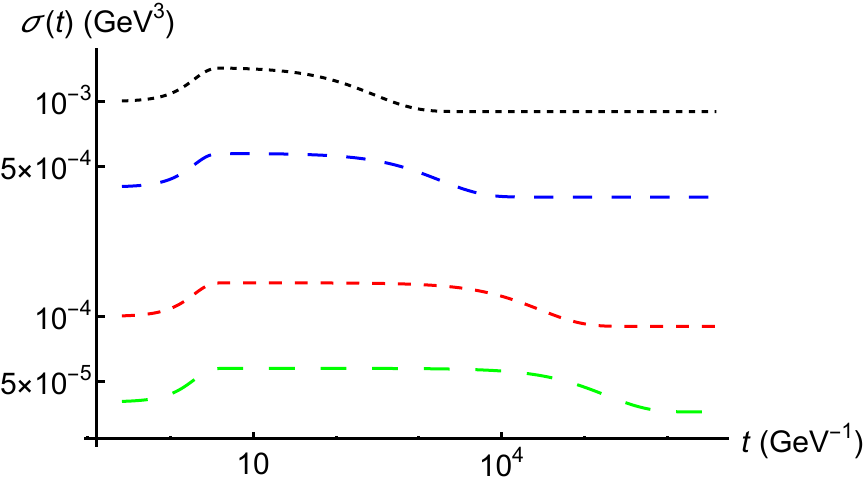}
        \put(85,50){\bf{(b)}}
    \end{overpic}
    \begin{overpic}[width=0.41\textwidth]{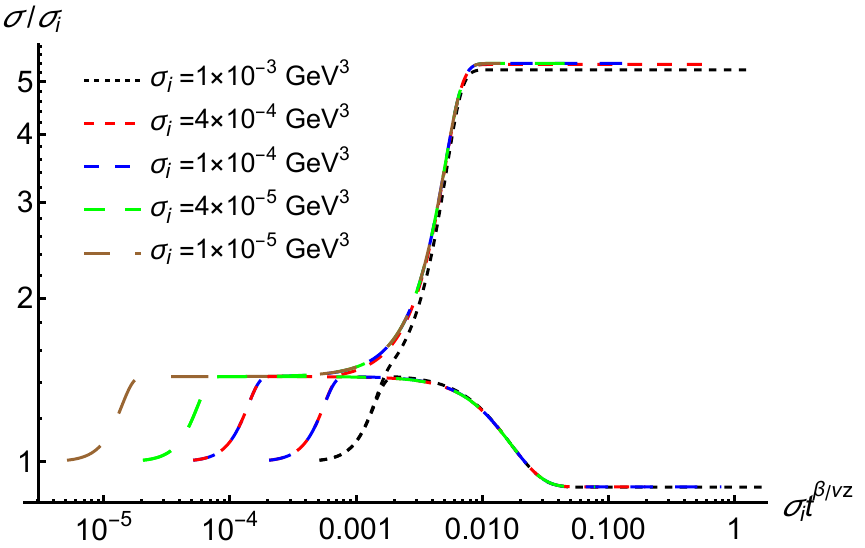}
        \put(87,50){\bf{(c)}}
    \end{overpic}
    \caption{Evolution curves of sigma condensate with \textbf{(a)} $-\epsilon\sigma_i^{-1/\beta}=3323$ and \textbf{(b)} $-\epsilon\sigma_i^{-1/\beta}=92$; \textbf{(c)} Scaling the evolution curves of sigma condensate in (a) and (b) based on Eq.~\eqref{scalingsigmarf}.}
    \label{fig:9}
\end{figure}
\begin{figure}[htbp]
    \centering
    \begin{overpic}[width=0.44\textwidth]{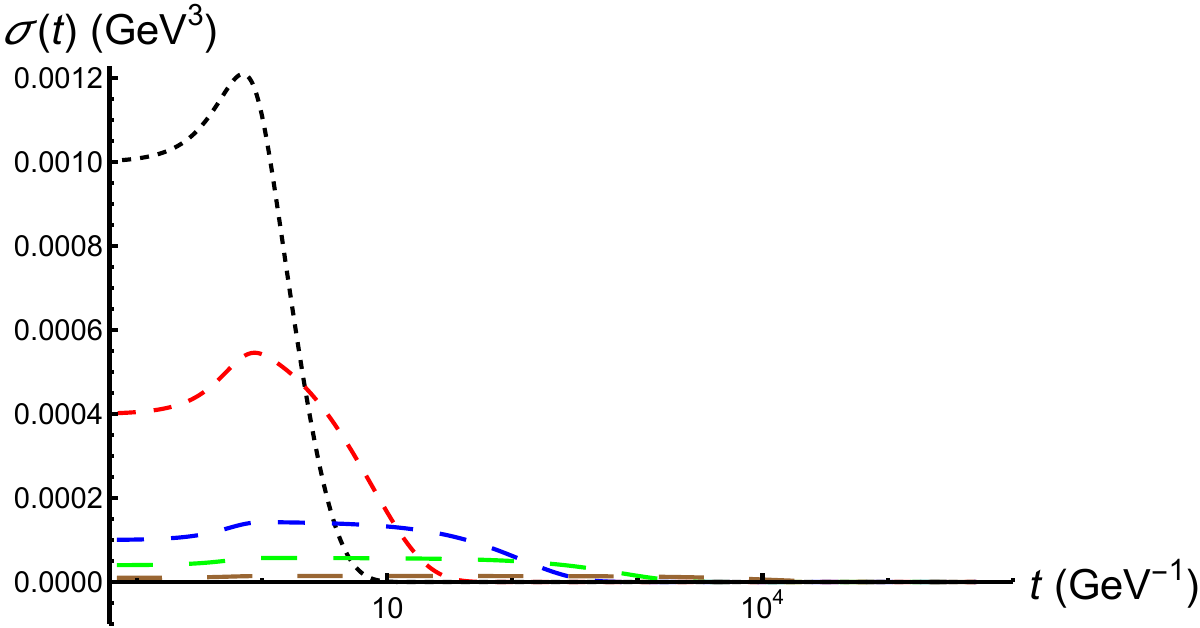}
    \put(85,43){\bf{(a)}}
    \end{overpic}
        \begin{overpic}[width=0.42\textwidth]{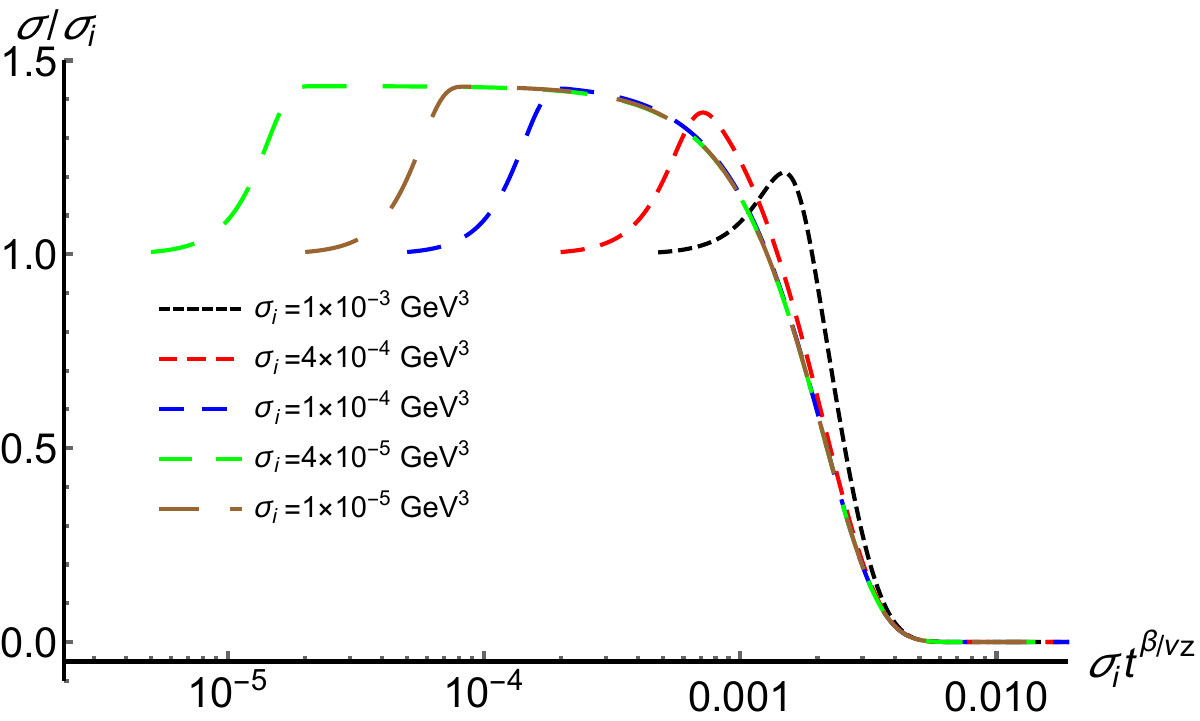}
    \put(85,50){\bf{(b)}}
    \end{overpic}
    \caption{\textbf{(a)} Evolution curves of sigma condensate with $\epsilon\sigma_i^{-1/\beta}=16677$; \textbf{(b)} Scaling the evolution curves of sigma condensate in (a) based on Eq.~\eqref{scalingsigmarf}.}
    \label{fig:10}
\end{figure}
In Fig.~\ref{fig:8}(a), there is obvious different on the order of the magnitude of initial $\sigma_i$, but the evolution curves share the same tendency and all can be separated into three stages. The intermediate stage is the period in which the short-time dynamic appears. Through numerically fitting the data according to Eq.~\eqref{scalingst} as shown in Fig.~\ref{fig:8}(b), we have $\theta\approx 0$ or $x\approx 1$. With the short-time dynamical critical exponent, the curves with different initial states can be completely scaled by Eq.~\eqref{scalingsigma} as shown in Fig.~\ref{fig:8}(c). Therefore, it is verified that the universal short-time behavior with the exponent $\theta=0$ arises in the perthermalization stage. It implicates that a smaller initial sigma condensate leads to a longer duration of the prethermalization stage.

If the final state is slightly deviation from the critical point, the evolution still can emerge the universal short-time behavior. Similar to the derivation of scaling form in Eq.~\eqref{scalingintermsoft}, one can derive the corresponding scaling form in terms of $t$ as
\begin{equation}
     \sigma(\sigma_i,\epsilon,m_q,t)=t^{-\beta/\nu z}f_{t} (\sigma_i t^{x\beta/\nu z},\epsilon t^{1/\nu z}, m_q t^{\beta\delta/\nu z}).
\end{equation}
One can also have the scaling form in terms of $\sigma_i$ as
\begin{equation}\label{scalingsigmarf}
    \sigma(\sigma_i,\epsilon,m_q,t)=\sigma_i^{1/x}f_{\sigma_i} (\epsilon \sigma_i^{-1/x\beta }, m_q \sigma_i^{-\delta/ x}, t \sigma_i^{\nu z/x \beta}).
\end{equation}
When the final state is $R_f=(T_f,0)$ with $\epsilon=T_f-T_c\neq 0$, the system will relax to the final equilibrium state with finite time at $\sigma\propto \epsilon^{\beta}$ in the ordered phase as shown in Fig.~\ref{fig:9}, or at $\sigma=0$ in the disordered phase as shown in Fig.~\ref{fig:10}. In Eq.~\eqref{scalingsigmarf}, the scaling function $f_{\sigma_i}$ has three variables. For simplicity, we use the projection method to analysis the multivariate scaling behavior. We get the evolution curves in Fig.~\ref{fig:9}(a) and (b) with fixing $-\epsilon \sigma_i^{-1/\beta}=3323$ and $92$, respectively. To verify the scaling function, based on Eq.~\eqref{scalingsigmarf}, we plot $\sigma(t)/\sigma_i^{1/x}$ versus $\sigma_i t^{\beta/\nu z}$ in Fig.~\ref{fig:9}(c). Since the short time dynamic exponent $\theta=0$, the curves overlap and behavior as a plateau in the prethermalization stage. In the long time limit, the curves overlap and show as horizontal lines at different values because of different fixing values. Note that the different fixing values only change the crossover position and have no impact on the short time scaling behavior.

\begin{figure}[htbp]
    \centering
    \begin{overpic}[width=0.46\textwidth]{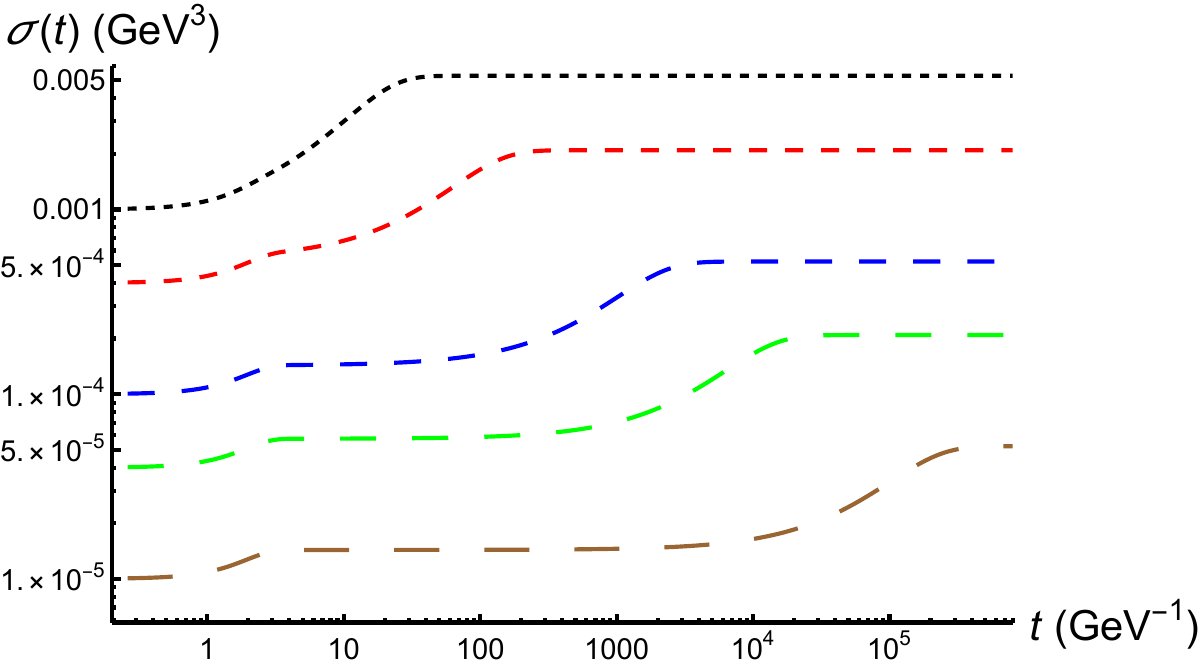}
    \put(85,50){\bf{(a)}}
    \end{overpic}
        \begin{overpic}[width=0.42\textwidth]{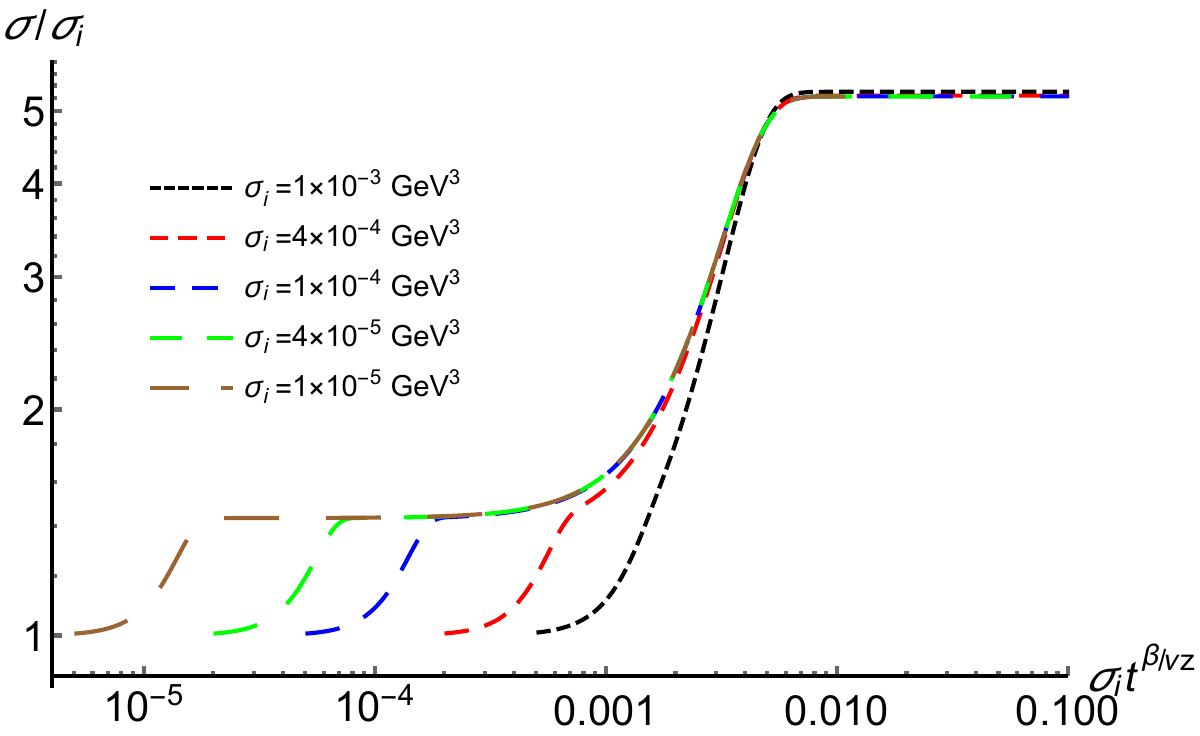}
    \put(85,48){\bf{(b)}}
    \end{overpic}
    \caption{\textbf{(a)} Evolution curves of sigma condensate with $ m_q \sigma_i^{-\delta}=3*10^6$; \textbf{(b)} Scaling the evolution curves of sigma condensate in (a) based on Eq.~\eqref{scalingsigmarf}.}
    \label{fig:11}
\end{figure}

In another case, we let the final state $R_f=(0,m_q)$ and fix $m_q \sigma_i^{-\delta}=3\times 10^{-6}$. In Fig.~\ref{fig:11}(a), the prethermalization stage arises after the microscopic-scale dominant region. Then the system crosses to the long-time thermalization stage. Finally, the system relaxes to the steady state and $\sigma\propto m_q^{1/\delta}$. Scaling the data according to the scaling form in Eq.~\eqref{scalingsigmarf}, we obtain overlapped curves in Fig.~\ref{fig:11}(b). These numerical analysis indicate that the scaling function $f_{\sigma_i}$ is well verified through the projection method.

\begin{figure}[htbp]
    \centering
    \includegraphics[width=0.48\textwidth]{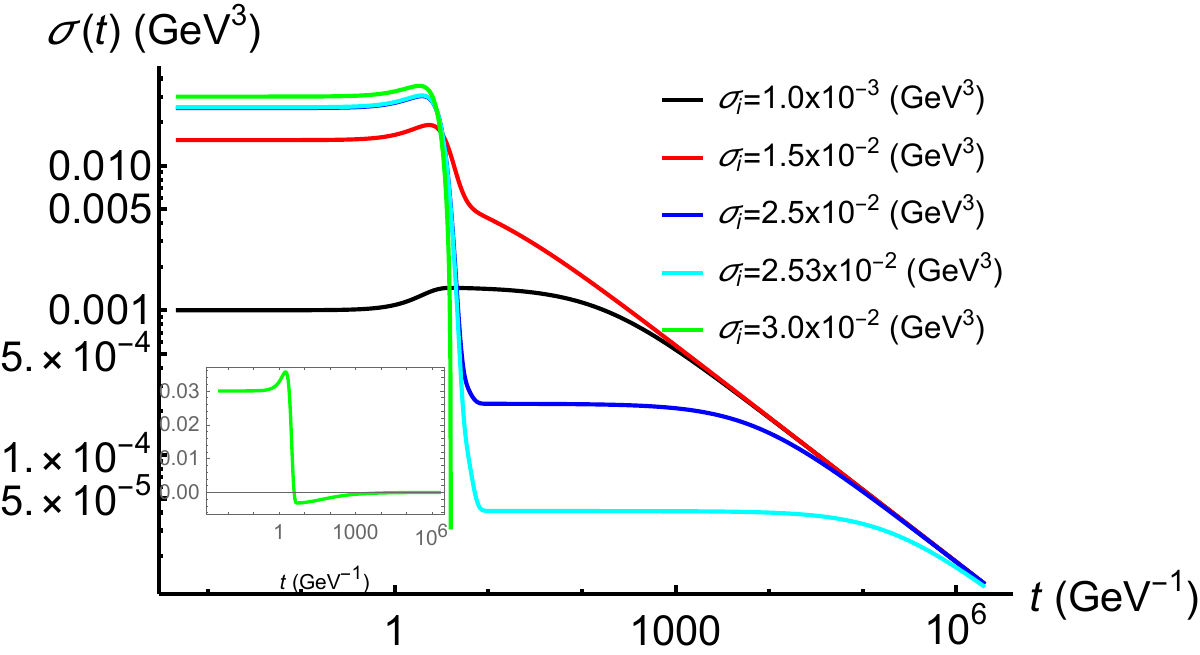}
    \caption{Initial state dependence of the sigma condensate evolution, $\sigma_i\gtrsim \sigma_{sat}$.}
    \label{fig:beyondsigmai}
\end{figure}

In above analyses, the short-time dynamic and the thermalization are realized with small initial sigma condensates. What if the magnitudes of the initial sigma condensate close to or larger than the saturated value? Generally, in small initial sigma condensate, before the universal short time dynamic, there are two particular properties in evolution. On the one hand, it is a non-universal period. On the other hand, this period is about in $0\leq t\leq t_{pre}\approx 2\pi T_f$. As shown in fig.~\ref{fig:beyondsigmai}, we find that when $\sigma_i$ is almost equals to or larger than $\sigma_{sat}$, $\sigma (t)$ show a steep decrease approximately at $t_{pre}$. If $\sigma_i\lesssim 2.54\times  10^{-2}\text{GeV}^3$, $\sigma(t)$ drops to a ``prethermalization'' state\footnote{It might be not a real prethermalization stage, but behaviors as a ``prethermalization''.}. However, if $\sigma_i$ are roughly larger than $2.54\times 10^{-2} \text{GeV}^3$, for example when $\sigma_i=3.0\times 10^{-2}\text{GeV}^3$, the sigma condensate drops to a negative value and approaches zero from the bottom, as shown in the inserted figure of fig.~\ref{fig:beyondsigmai}.

\subsubsection{Quench from the ordered phase}

As the schematic program shown in Fig.~\ref{sketch}, we will also study the short-time dynamics from $A'\rightarrow C$ and $B\rightarrow C$ . Without considering the high orders, the five dimension scalar field behaves as $\chi(r) \propto r^3$ in the critical region. It is interesting to explore that whether  the scaling behavior in this case is the same as the case of $A\rightarrow C$.

\begin{figure}[htbp]
    \centering
    \begin{overpic}[width=0.48\textwidth]{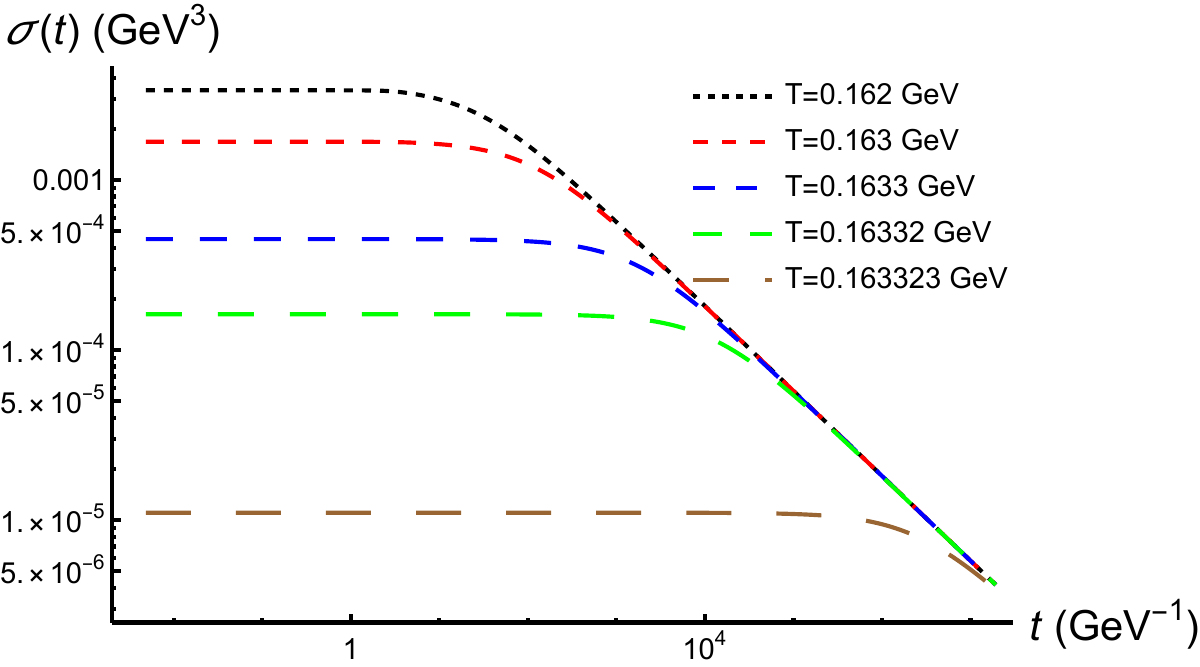}
    \put(83,50){\bf{(a)}}
    \end{overpic}
        \begin{overpic}[width=0.45\textwidth]{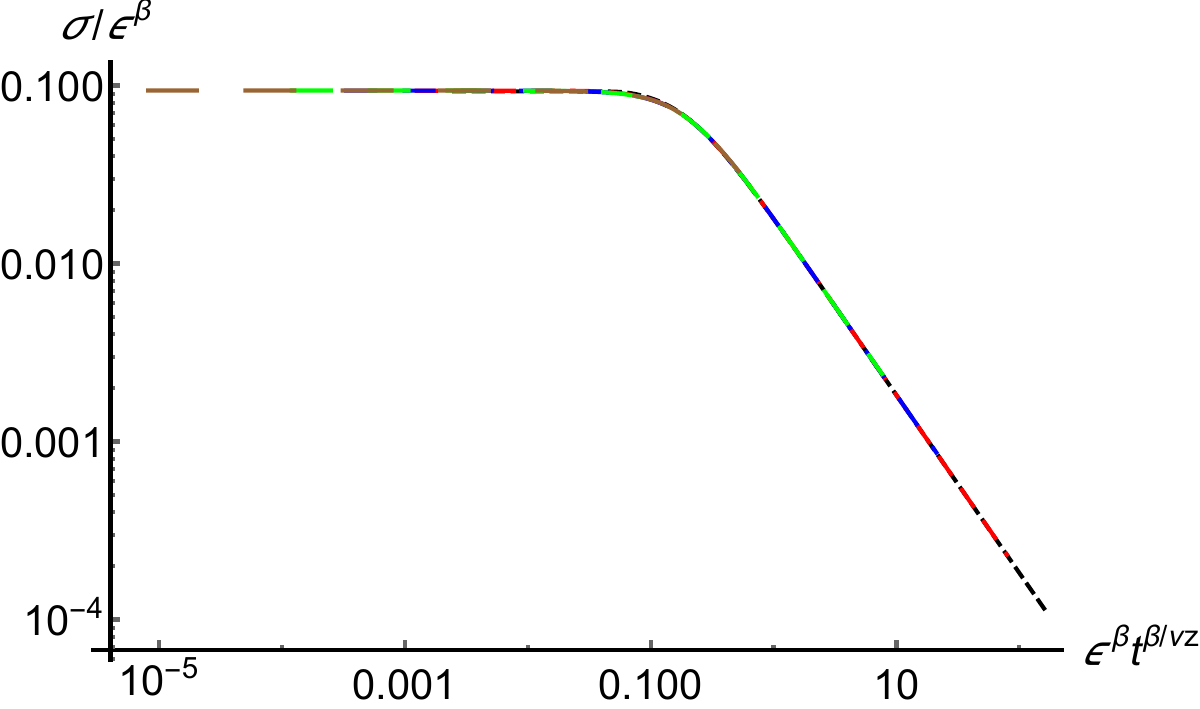}
    \put(85,48){\bf{(b)}}
    \end{overpic}
    \caption{\textbf{(a)} Evolution curves of sigma condensate with different initial temperature; \textbf{(b)} Scaling the evolution curves of sigma condensate in (a) based on Eq.~\eqref{scalingepsilonmi}(a)}
    \label{fig:13}
\end{figure}

\begin{figure}[htbp]
    \centering
    \begin{overpic}[width=0.46\textwidth]{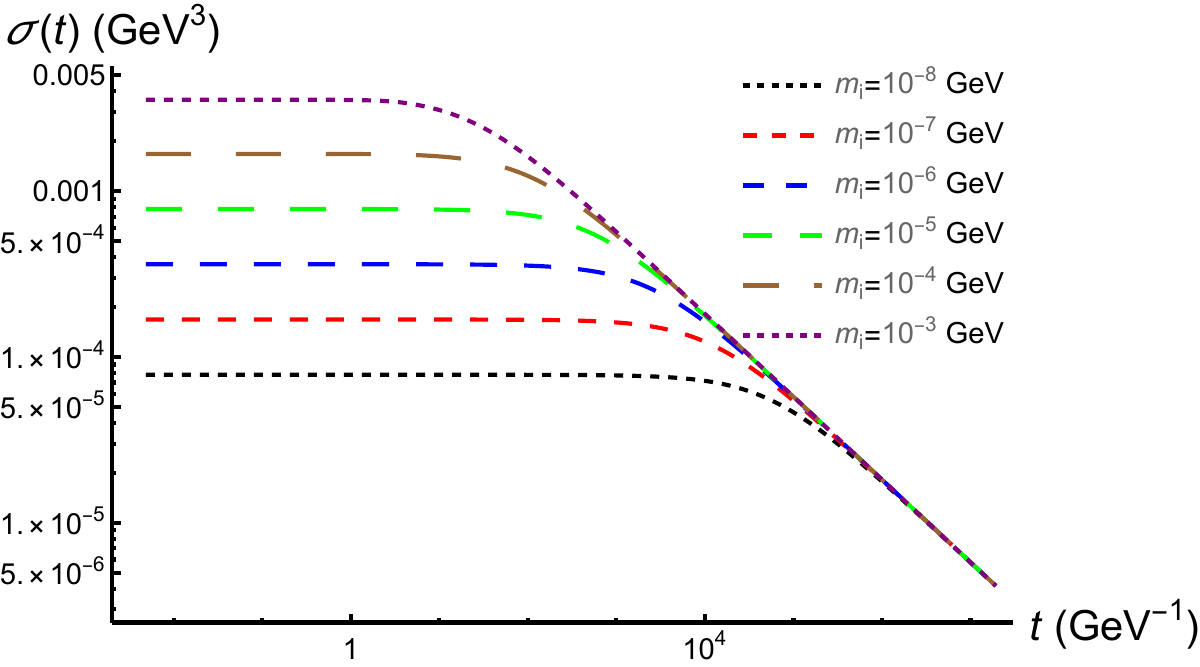}
    \put(85,52){\bf{(a)}}
    \end{overpic}
        \begin{overpic}[width=0.48\textwidth]{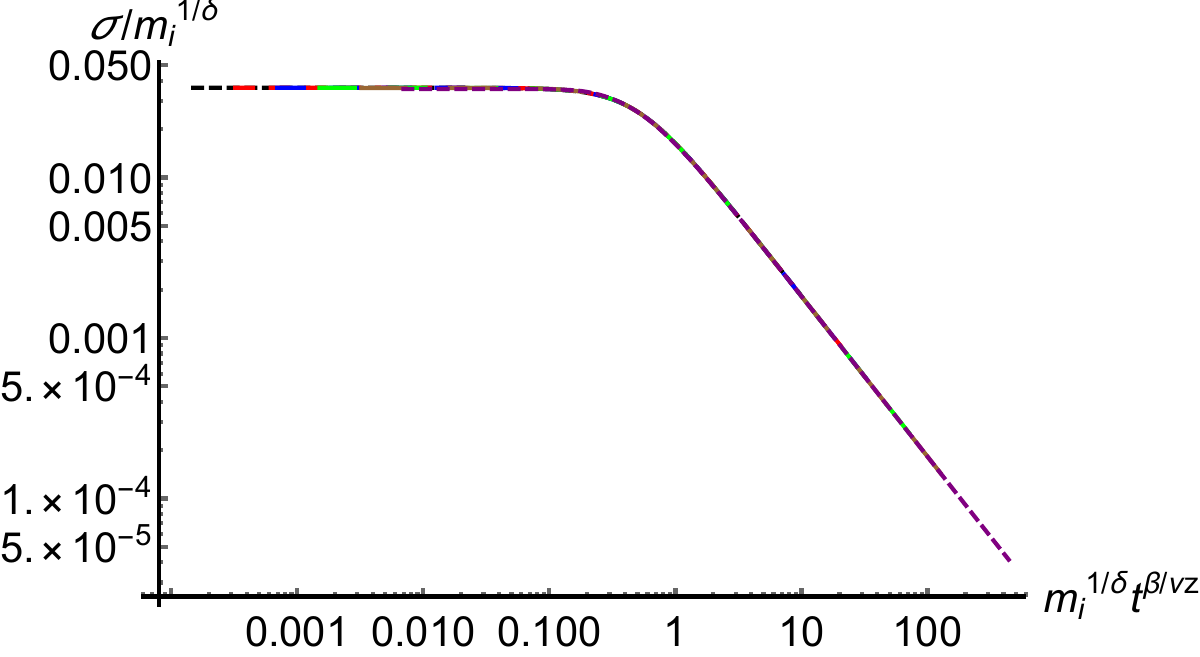}
    \put(85,50){\bf{(b)}}
    \end{overpic}
    \caption{\textbf{(a)} Evolution curves of sigma condensate with different initial quark mass; \textbf{(b)} Scaling the evolution curves of sigma condensate in (a) based on Eq.~\eqref{scalingepsilonmi}(b)}
    \label{fig:14}
\end{figure}

One can apply the initial state at $R_i=(\epsilon_i, m_i)$ and sudden quench the system to the critical point. By choosing the particular scaling, the scaling form of Eq.~\eqref{scalingepsilon} transform to
\begin{equation}
    \sigma(\epsilon_i, m_i,t)=t^{-\beta/\nu z} f_{t}(\epsilon_i t^{x/\nu z},m_i t^{x\beta\delta /\nu z}),
\end{equation}
and
\begin{subequations}\label{scalingepsilonmi}
\begin{eqnarray}
    \sigma(\epsilon_i,m_i,t)&=&\epsilon_i^{\beta/x}f_{\epsilon_i}(m_i\epsilon^{-\beta\delta}, t\epsilon_i^{\nu z/x}),\\
    \sigma(\epsilon_i,m_i,t)&=& m_i^{1/\delta x} f_{m_i}(\epsilon_i m_i^{- 1/\beta \delta}, t m_i^{\nu z/x\beta\delta}).
\end{eqnarray}
\end{subequations}
Due to the complexity of considering finite $\epsilon_i$ and $m_i$ simultaneously, we separately consider either finite $\epsilon_i$ or $m_i$ at once. When $R_i=(\epsilon_i,0)$, $\sigma(t)$ should behave as $\sigma\propto \epsilon_i^{\beta}$ in the prethermalization stage, so that $f_{t}\propto (\epsilon_i t^{x/\nu z})^{\beta}$. Thus, one has the leading scaling term as
\begin{equation}
    \sigma\propto \epsilon_i^{\beta} t^{(x-1)\beta/\nu z}=\epsilon_i^{\beta}t^\theta.
\end{equation}
In the long-time region, the scaling function $f_{t}$ must reduce to a constant, so that the evolution of the sigma condensate is reduced to the critical slowing down scaling $\sigma\propto t^{-\beta/\nu z}$, which has been studied in Sec.~\ref{ctsd}. The numerical results are shown in Fig.~\ref{fig:13}(a). The evolution has two different features, the prethermalization, and the thermalization stages. In Fig.~\ref{fig:13}(b), according to Eq.~\eqref{scalingepsilonmi}(a), by scaling the $\sigma(t)$ with $\epsilon_i^{\beta}$, the curves of $\sigma(t)/\epsilon_i^\beta$ overlap as a function of $\epsilon_i^\beta t^{\beta/\nu z}$.


Similarly, we let $R_i=(0,m_i)$ and study the scaling function in Eq.~\eqref{scalingepsilonmi}(b). Along with the evolution, firstly, sigma condensate behaviors as $\sigma(t)\propto m_i^{1/\delta}t^{(x-1)\beta/\nu z}=m_i^{1/\delta}t^\theta$ in the prethermalization stage, then crosses to $\sigma(t)\propto t^{-\beta/\nu z}$ in the thermalization stage. These scaling predictions are verified in Fig.~\ref{fig:14}. In Fig.~\ref{fig:13}(a), it is the original data for $\sigma(t)$. The curves have the prethermalization and the thermalization stages and their crossover. As predicted by the scaling form in Eq.~\eqref{scalingepsilonmi}(b), the $\sigma(t)/m_i^{1/\delta}$ versus $m_i^{1/\delta}t^{\beta/\nu z}$ curves overlap each other in Fig.~\ref{fig:14}.

\begin{figure}[htbp]
    \centering
    \begin{overpic}[width=0.48\textwidth]{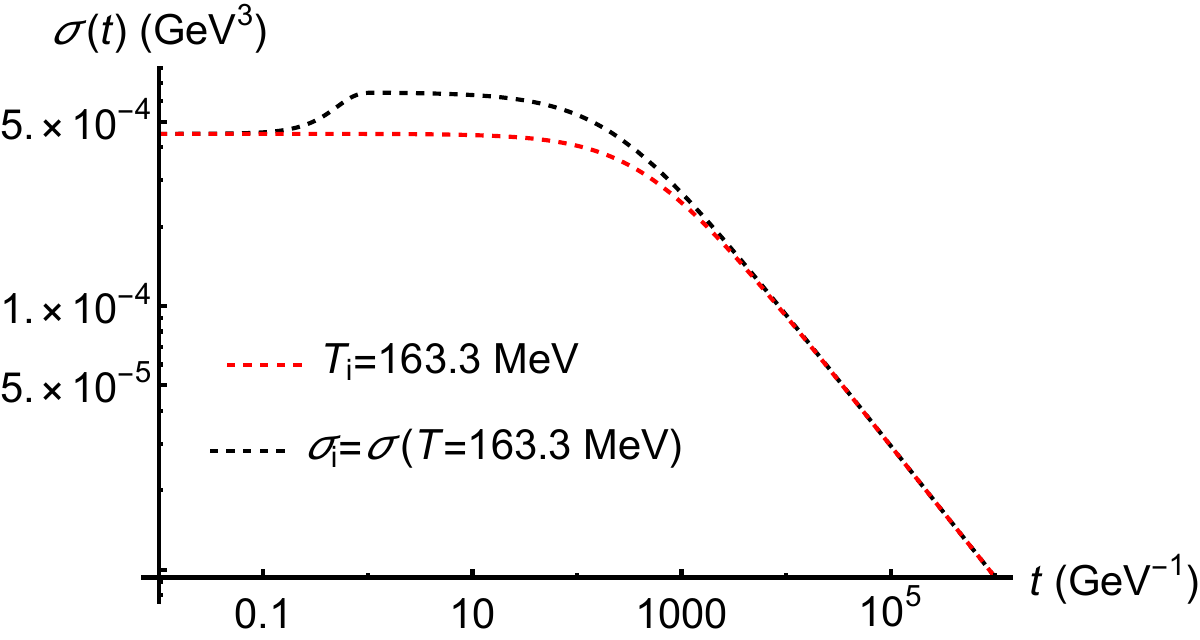}
    \put(83,48){\bf{(a)}}
    \end{overpic}
        \begin{overpic}[width=0.45\textwidth]{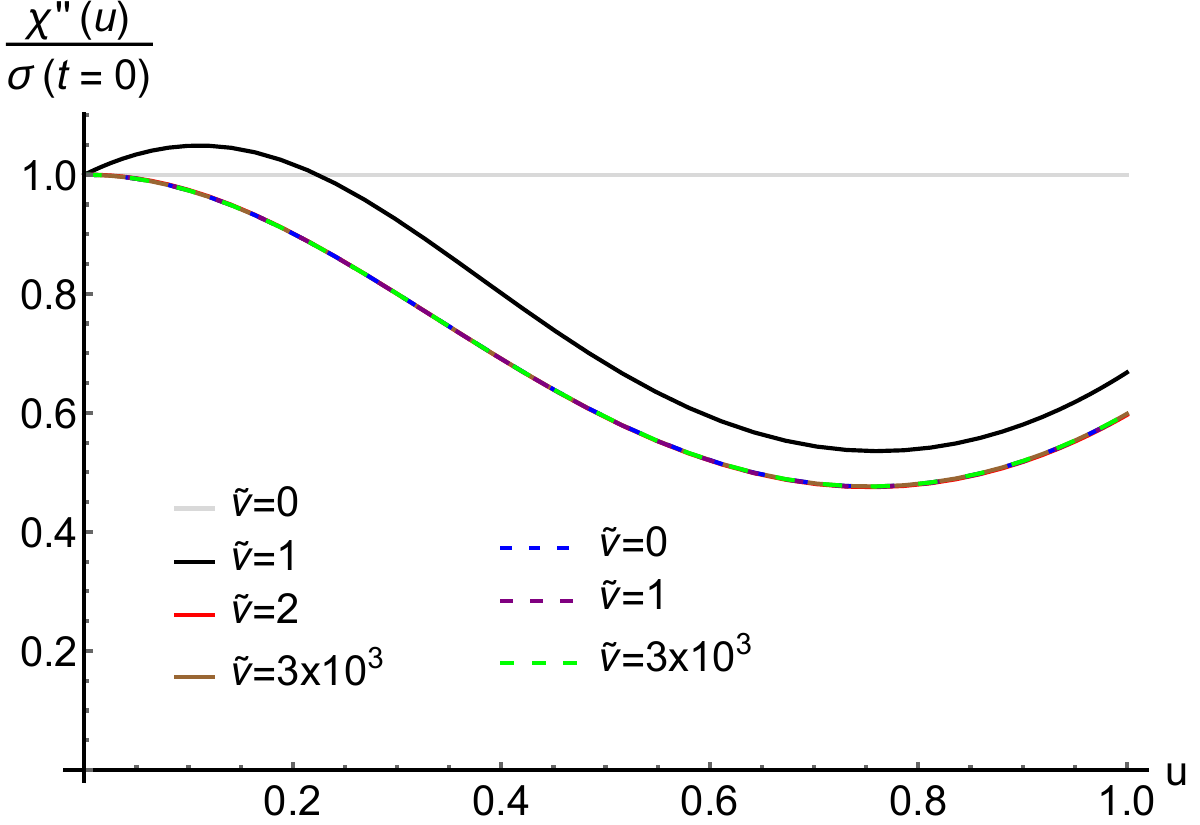}
    \put(85,50){\bf{(b)}}
    \end{overpic}
    \caption{\textbf{(a)} Time dependence of $\sigma(t)$ with equal $\sigma_i$ through different quenching protocols, $A\rightarrow C$ and $A'\rightarrow C$; \textbf{(b)} Curves of $\chi''(u)/\sigma(t=0)$ at different $\tilde{\nu}$.}
    \label{fig:15}
\end{figure}

It seems that the evolution in Figs.~\ref{fig:13} and~\ref{fig:14} do not have obvious boundary between the prescaling and prethermalization stages, which are primarily different from fig.~\ref{fig:8}. To analysis the difference between them, we compare the evaluations of  sigma condensate $\sigma(t)$ and the second derivation of chiral field with respect to $r$ $\chi''(r)$. For convenience of numerical calculation, we have a coordinate transform in Eq.~\eqref{eom-chi},
\begin{equation}
u={r}/{r_h},\ \ \ \ \ \ \tilde{v}=v/r_h.
\end{equation}
In Fig.\ref{fig:15}, we show the numerical results of $\sigma(t)$ and
$\chi''(u)/\sigma(t=0)$ obtained from the quench paths $A\rightarrow C$ and $A'\rightarrow C$ with equal initial sigma condensate in Fig.~\ref{fig:14}. The sigma condensation increases in the very beginning non-universal period in the case of $A\rightarrow C$. However, the sigma condensation almost keeps invariant in the non-universal period in the case of $A'\rightarrow C$. From the aspect of $\chi''(u)/\sigma(t=0)$, in the prescaling stage, in case $A\rightarrow C$, $\chi''(u)/\sigma(t=0)$ has a changing process from a constant to a time independent function. In case $A'\rightarrow C$, $\chi''(u)/\sigma(t=0)$ always a time independent function.

It is interesting that the ratio is stable in the prethermalization stage and equal to the values in the thermalization stage. It means that the time-dependent term and $u$-dependent part are decoupled in Eq.~\eqref{eom-chi}. To some extent, even though the scaling form in prethermalization is totally different from the thermalization, there are some particular physical quantities which behavior like the (quasi-)equilibrium state.
Furthermore, as a result of the decoupling of $t$ and $u$, if the sigma condensate satisfy the power-law scaling $\sigma(t)\propto t^{-a}$ with $a>0$, i.e., $\sigma(\tilde{v})\propto \tilde{v}^{-a}$ for $t=\tilde{v}\pi T$ at $r=0$, one has $\chi(\tilde{v},u)=\sigma(\tilde{v})g(u)$. From Eq.~\eqref{eom-chi-rescale}, one can derive $-a-1=-3a$, thus $a=1/2$, with dimension analysis. Moreover, there is another solution $\sigma(t)\propto t^0$. These two solutions correspond to the thermalization and prethermalization, respectively.

\section{Conclusions and discussions}\label{sec:summary}
The soft-wall AdS/QCD model provides an effective holographic framework to deal with the nonperturbative problems of QCD, especially for chiral symmetry breaking and restoration. Since the model contains information of the order parameter, it would be quite interesting to extend the previous equilibrium studies to nonequilibrium phase transition.

By quenching the system from initial states deviating from the equilibrium states, we solve the real-time evolution of chiral condensate in the two-flavor ($N_f=2$) soft-wall AdS/QCD model. At this stage, we work in the probe limit, i.e. neglecting the back-reaction to the background geometry. In this way, we are considering the thermalization of the system under an infinite heat bath.

It is shown that, at very low temperatures, the chiral condensate shows oscillating behaviors while its amplitude decay exponentially with time. At higher temperatures but still below $T_c$, the oscillation disappears and only the fast exponential damping is left. We compare the oscillating frequencies and the relaxation times with the complex frequency of quasi-normal modes. It is found that they match with each other very well. Therefore, the late-time thermalization of the system could be described by the quasinormal modes. Furthermore, it is also found that the relaxation time would diverge when the temperature of the heat bath approaches $T_c$, showing a typical behavior of critical slowing down. The exponential damping turns to a power law, and by fitting the late time behavior, we get the dynamical critical exponent $z\approx2.0$.

Besides the late time thermalization, it is more interesting to observe that, starting from a large class of initial states, the system would linger over a quasi-steady state for a certain period of time before the thermalization, which is very similar to the interesting phenomenon named prethermalization in QCD community. According to the extracted initial-slip exponent $\theta\approx0$, it is observed that such quantity is still a mean-field value due to the large $N$ suppression. Therefore, there are some important open issues that can be studied in the future. From a theoretical point of view it is crucial to explore how to vary the initial conditions to realize the scenario of the non-thermal fix point in the soft-wall AdS/QCD model, and that whether the collective modes of two out-of-equilibrium universality, characterized by the NTFP and the initial-slip exponent, are the same.

In the prethermalized regime, $\theta$ arises as a new universal critical exponent, which forms as the memory of initial configurations. Taking into account the rapidly decayed magnetic field which is produced in the off-central heavy ion collisions, the mass of charged particles, such as protons, is strongly altered by the magnetized environment, which presents as a unique scenario to study the short-time dynamical scaling and its subsequent behaviors~\cite{Xu:2020sui}. A prolonged stage also affects the estimation of rapidity-distribution of the charged particles as a function of collision centrality in experimental measurements. A quantitative impact of the initial-slip exponent on the two- and higher correlation functions of protons remains to be evaluated.

Triggered by the forthcoming experimental measurements at high baryon chemical potentials, the description of fireballs created in low-energy collisions need to be contained the phase transition of first order~\cite{STAR:2020dav}. The corresponding time evolution has to be improved in two aspects. Firstly, the soft-wall AdS/QCD model should extend to finite chemical potentials, where the first order phase transition occurs and the phase boundary beyond the critical point is located~\cite{Critelli:2017euk}. Secondly, coupling with the baryon density, the chiral condensate is governed by the conservation law and its linear-response is presented by the diffusive equation. The influence of the short-time dynamical behaviors on the possible spinodal instability and other associated phenomena are under way and will be published elsewhere.

\vspace*{1cm}
{\bf Acknowledgements}\quad
We would like to thank the useful discussion with  Yidian Chen, Song He, Mei Huang, Lang Yu and Xinyang Wang. X.C. is supported by the National Natural Science Foundation of China under Grant No. 12275108 and the Fundamental Research Funds for the Central Universities under grant No. 21622324.  J.C. is supported by the start-up funding from Jiangxi Normal University under grant No. 12021211. H.L. is supported by the National Natural Science Foundation of China under Grant No. 11405074. D.L. is supported by the National Natural Science Foundation of China under Grant Nos. 12275108, 12235016, 11805084, the PhD Start-up Fund of Natural Science Foundation of Guangdong Province under Grant No. 2018030310457 and Guangdong Pearl River Talents Plan under Grant No. 2017GC010480.

\vspace*{1cm}
\appendix
\section{Pseudospectral method}\label{appendixA}

The pseudospectral method~\cite{boyd2001chebyshev,hesthaven2007spectral} is a very effective numerical method with high accuracy for solving the initial-value problem of the partial differential equation. The following is a brief overview to simulate the EOM in Eq.~\eqref{eom-chi-rescale} by the pseudospectral method.

Firstly, we apply the variable substitution, $v\rightarrow \tilde{v}\equiv v/r_h$ and $r\rightarrow u\equiv r/r_h$. Then Eq.~\eqref{eom-chi} becomes
\begin{widetext}
\begin{eqnarray}\label{eom-chi-rescale}
2\partial_{\tilde{v}}\partial_u \chi(\tilde{v},u)-\left[\frac{3}{u}+\Phi '(u)\right]\partial_{\tilde{v}} \chi (\tilde{v}, u)-f(u)\partial_u^2\chi(\tilde{v},u)&\nonumber\\
+\left[\frac{3}{u} f(u)+\Phi'(u)f(u)-f'(u)\right]\partial_u\chi(\tilde{v},u)+\frac{1}{ u^2}(m_5^2+\frac{\lambda}{2}\chi(\tilde{v},u)^2)\chi (\tilde{v},u)&=0.
\end{eqnarray}
\end{widetext}

Generally, the function $\chi(r)$ can be expanded into the nodal expansion,
\begin{equation}
    \chi(u)=\sum_{i=0}^{N} \chi_i l_i(u)
\end{equation}
with
\begin{eqnarray}
   \chi_i &=&\chi(u_i)\nonumber\\
   l_i(u) &=&\mathop{\Pi}\limits_{i=0,i\neq j}^{N}\frac{u-u_i}{u_j-u_i},\nonumber
\end{eqnarray}
where $\{u_i\}_{i=0}^N$ are the collocation points and $\{l_i\}_{i=0}^N$ are the basis functions. In this expansion, the undetermined parameters are directly the function values at the collocation points. For the optimal scenario, the collocation points or the grid points $\{u_i\}_{i=0}^N$ for the basis are given by the Chebyshev-Gauss-Lobatto points~\cite{BALTENSPERGER199941}.
The discrete $u$ in the interval $[0,1]$ is
\begin{eqnarray}
u_i=\frac{1}{2}\left [1-\cos \left(\frac{i-1}{N-1}\pi\right)\right],
\end{eqnarray}
with $j=1,2,\cdots , N$. In this work, we choose $N=60$. The derivative operator $\partial _x$ is approximately replaced by a discrete finite difference derivative $\hat{D}$. At the point $u_j$, one obtains the $p$-order derivative as
\begin{equation}
    \chi^{(p)}(u_j)=\hat{D}^{(p)}\chi(u_j)=\sum_{i=0}^{N}\chi_i l_i^{(p)}(u_j)
\end{equation}

In our calculation, the derivation is realized by employing the build-in ``FiniteDifferenceDerivative'' operator in Mathematica. Then, we get a series of equations in the form of
\begin{eqnarray}\label{EOM-numerical}
\partial_{\tilde{v}} \chi(\tilde{v},u_i)= F[\hat{D},u_i,\chi(\tilde{v}, u_i)]
\end{eqnarray}
with $i=1,2,\cdots, N$. Thus, a second-order partial differential equation is transformed into the first-order ordinary differential equation. With prepared initial conditions $\chi(0,u_i)=\chi_0(u_i)$, it is straightforward that these equations are solved.

\bibliographystyle{apsrev4-2}
\bibliography{prethermal}
\end{document}